\documentclass{aa501}
\usepackage{graphicx}
\usepackage{psfig}
\usepackage{amsmath}
\usepackage{amssymb}
\renewcommand{\baselinestretch}{1.0}

\newcommand{\kms}{km~s$^{-1}$}
\newcounter{qub}
\newcommand{\qq}{\addtocounter{qub}{1}\arabic{qub}}

\begin{document}

\title{Studies of galaxies in voids}
\subtitle{I. \ion{H}{i} observations of Blue Compact Galaxies}

\author{S.A.~Pustilnik\inst{1,6} \and J.-M.~Martin\inst{2} \and
W.K.~Huchtmeier\inst{3} \and N.~Brosch\inst{4}
\and V.A.~Lipovetsky\inst{1}\thanks{Deceased 1996 September 22.}
\and G.M.~Richter\inst{5} }

\offprints{S. Pustilnik \email{sap@sao.ru}}

\institute{Special Astrophysical Observatory RAS, Nizhnij Arkhyz,
Karachai-Circassia,  369167 Russia
\and
Observatoire de Paris-Meudon, Meudon, France
\and
Max-Planck-Institut f\"{u}r Radioastronomie, 53121 Bonn, Germany
\and
Wise Observatory, Tel-Aviv University, Tel-Aviv 69978, Israel
\and
Astrophysikalisches Institut Potsdam, An der Sternwarte 16, D-14482 Potsdam,
Germany
\and
Isaac Newton Institute of Chile, SAO Branch
}

\date{Received May 7, 2001; accepted  April 16, 2002}

\abstract{
We present here results of studies of the properties of galaxies located in
 very low  density environments. We observed 26 blue compact galaxies
(BCGs) from the  Second Byurakan (SBS) and Case surveys located in voids
with the radial velocities $V_\mathrm{hel} \lesssim$11000~\kms, two BCGs in the
void behind the Virgo cluster and 11 BCGs in denser environments.
\ion{H}{i} fluxes and profile widths, as well as estimates of total \ion{H}{i}
masses, are presented for the  27 detected galaxies (of which 6 are
in three galaxy pairs and are not resolved by the radiotelescope beam).
Preliminary comparisons of void BCGs with similar objects from intermediate
density regions - in the general field and the Local Supercluster
(sub-samples of BCGs in the SBS zone) and in the dense environment of the
Virgo Cluster (a BCD sample) -  are performed using the
hydrogen-to-blue-luminosity ratio $M(\ion{H}{i})$/$L_\mathrm{B}$. We find  that
for
the same blue luminosity, for $M_\mathrm{B} > -$18\fm0, BCGs in lower density
environment have on average  more \ion{H}{i}. The slope $\beta$
of the $M(\ion{H}{i})$/$L_\mathrm{B} \varpropto L^{\beta}$ for BCGs shows a trend
of steepening with decreasing bright galaxy density, being very close
to zero for the densest environment considered here and reaching
$\beta$= --0.4 for voids.
}

\authorrunning{S.A. Pustilnik et al.}

\titlerunning{Studies of galaxies in voids. I.}
  \maketitle


\section{Introduction}

Simple models for structure formation in CDM cosmologies predict
a detectable biasing
in the relative space distributions of bright  (luminous, massive)
vs. faint (subluminous, dwarf) galaxies, as a consequence of gaussian peak
correlations in the random field (Bardeen et al.~\cite{Bardeen86}, Dekel
\& Silk~\cite{Dekel86}).
Thus, large regions devoid of massive galaxies (voids) emerge as a natural
consequence of this approach. Low-mass and underdeveloped galaxies were
expected to populate  voids.
 Simple CDM models were shown to be too simplistic to
 explain the entire range of relevant observational data (e.g.,
 Ostriker~\cite{Ostriker96}). More complicated versions of DM cosmology were
 suggested, such as Cold+Hot DM (e.g., Klypin et al.~\cite{Klypin93}) or
unstable  DM (e.g., Doroshkevich et al.~\cite{Doroshkevich89}, Cen
\cite{Cen01}).

The observational evidence for the presence of galaxies in
voids has been somewhat controversial.
No large population of faint galaxies, with a much more homogeneous
distribution than that of bright ones, was detected. However, the conclusions
on how closely do dwarf galaxies follow the large scale structure
(LSS) delineated by bright
galaxies were quite different. Thuan et al.~(\cite{Thuan91}, and references
therein) claimed that low surface brightness (LSB) dwarfs are distributed
just like bright galaxies in volumes with limiting radial velocities
up to 3000 \kms.
On the other hand, Salzer~(\cite{Salzer89}), Pustilnik et
al.~(\cite{Pustilnik95}), and Lee et al.~(\cite{Lee00}), who studied
H{\sc ii}-galaxies (BCGs) with radial velocities up to
10000 \kms, found statistically significant
differences in the spatial distributions of these low luminosity (low mass)
and bright (massive) galaxies.  Similar results are obtained by
Popescu et al.~(\cite{Popescu97a}) in their Heidelberg void survey.
While the majority of BCGs follow
the filaments delineated  by bright galaxies, being significantly more
scattered around these  structures,
some 10--15 per cent of BCG/H{\sc ii} galaxies were found to populate the
voids seen in the distribution of bright galaxies ($M_\mathrm{B}$ $<$ --19.5,
$H_\mathrm{0}$ =
75 \kms~Mpc$^{-1}$ throughout the paper). Furthermore, the
consideration of the nearest-neighbour statistics applied to the
low-surface brightness dwarf  sample
from Thuan et al.~(\cite{Thuan91}) leads to the conclusion that their spatial
distribution is similar to that of BCGs (Pustilnik et al.~\cite{Pustilnik95}).
Later analysis of this LSB galaxy sample by Dominguez-Tenreiro et al.
(\cite{Dominguez96}) yielded similar results.

  Most \ion{H}{ii} galaxies are subluminous, low-mass galaxies, which would be
very difficult to detect in large voids far outside the Local Supercluster
if they were not in a phase of active star formation (SF). Thus, with some
reservations, we can consider BCG/\ion{H}{ii} galaxies
as representative low-mass gas-rich galaxies and study their spatial
distribution  at much larger distances. Another important aspect of BCGs in
voids is that they can indicate where inside
voids one should search for galaxies with lower levels of SF activity.

Finally, due to the existence of super-large scale structure,
the voids themselves can have quite different properties such as
the underdensity value (Pustilnik et al.~\cite{Pustilnik94}).
Therefore the study of galaxies in voids yields important information for
the voids that host them.

Peebles~(\cite{Peebles01}) argued recently that the void phenomenon remains
the crucial one to reconcile the predictions of CDM models of structure
formation (including the most up-dated $\Lambda$CDM-versions) with
observations of the galaxy distribution.
While most predictions of CDM models seem consistent with
observations, the void galaxies do pose some problems.
The current simulations predict too many galaxies in voids. Either reasonable
mechanisms for the suppression of galaxy formation in voids should be involved
(Cen \& Ostriker~\cite{Cen00}), or the inclusion of warm dark matter
might help (Bode et al.~\cite{Bode01}).

An important question about galaxies populating voids is how similar
are they
to the more common galaxy population. The observed properties of  galaxies
as they appear today are determined partly by their formation parameters,
as outlined, e.g., by Dalcanton et al. (\cite{Dalcanton97}), and partly by the
influence of other galaxies through various interactions (e.g., in "Physics of
Nearby Galaxies" ed. by Thuan et al.~\cite{Thuan92}).
Galaxies in voids, by their apparently extreme isolation (at least from
luminous galaxies), could be thought of as being frozen in their nascent state
and could, in principle, reflect the distribution of global parameters of
the first-formed galaxies.
This can be very important, since in the framework of the
hierarchical galaxy formation (e.g., White \& Frenk~\cite{White91}), most
galaxies in more common environments have probably experienced some
transformations, no longer representing  the original parameters of a
primordial galaxy population.

One of the approaches to the large-scale structure
advocates a mass hierarchy in the LSS (e.g., Lindner et
al.~\cite{Lindner95,Lindner96}).
This assumes that larger mass objects determine larger linear
scale structures, both for overdense and for underdense regions.
In this interpretation, void BCGs, as representatives of low-mass
population, should belong to some overdensity structures defined by
filaments of fainter galaxies (with lower masses and luminosities) that are
mostly missed by modern redshift surveys dealing with
magnitude-limited samples.

In this paper we  study the properties of a large group of
low-mass, actively star-forming galaxies residing in voids.

Previous studies of void galaxy properties include those in the
Bootes void (e.g., Szomoru et al.~\cite{Szomoru96a,Szomoru96b}; Weistrop
\cite{Weistrop94}; Sage et al.~\cite{Sage97};
Cruzen et al.~\cite{Cruzen97}).
These galaxies represent a relatively high-luminosity group (with
$M_\mathrm{B}$ $<$ --21.0) and probably form an overdense region inside the
Bootes void.
Most of them are emission-line galaxies (ELGs) of various types of activity
-- from Starburst Nuclei (SBN) to Seyfert~2 galaxies.

One of the detailed, but somewhat limited (on galaxy number) studies of void
low-mass
galaxies is the Heidelberg void survey (Hopp et al.~\cite{Hopp95}, Popescu et
al. \cite{Popescu96,Popescu97a,Popescu97b,Popescu98}, Hopp \cite{Hopp98}).
In the Hopp et al. sample in the direction of voids only nine  dwarf
galaxies are located fully within voids (three objects), or
in some intermediate regions near the void boundaries (six galaxies). These
galaxies were observed in the \ion{H}{i} 21 cm line
by Huchtmeier et al.~(\cite{Huchtmeier97}) (hereafter HHK) and their
morphology and colors were studied by Vennik et al. (\cite{Vennik96}).
In the sample of Popescu et al. there were also 17 void BCGs
(eight fully located in voids), but these were not observed in \ion{H}{i}.
The conclusion of Huchtmeier et al.~(\cite{Huchtmeier97}) study was that the
ratio $M(\ion{H}{i})$/$L_\mathrm{B}$ for the nine
most isolated dwarfs is a factor of three larger than for a sample
of Virgo cluster dwarfs.

The above-mentioned results offer a first insight into the
properties of low-mass galaxies in voids, but with very poor statistics.
Other factors can presumably affect the comparison of different
samples.
In particular, the ratio $M(\ion{H}{i})$/$L_\mathrm{B}$ is the same, within the
uncertainties,
for the two HHK groups with nine dwarf galaxies each, populating
low and high galaxy density regions.

An additional study of galaxy properties in regions of lower galaxy density
was presented recently by Grogin \& Geller~(\cite{Grogin00a,Grogin00b}).
It is based on a large sample, mainly within the magnitude limits of the
Zwicky catalog, implying rather luminous galaxies for the range of
radial velocities studied  (5000 to 10000 \kms). However,
a significant number of fainter galaxies was added as well. Only a small
fraction of galaxies of their LV sample
reside in the type of environment discussed below for void BCGs.

In this paper we present the results of \ion{H}{i} observations of void
BCG/\ion{H}{ii} galaxies and  test for possible differences in their neutral
gas content through a
comparison with galaxies in the denser environments.
The paper is organized as follows: in Sect.~\ref{section:sample} we
describe the void sample. Sect.~\ref{section:obs} presents the observations
and data reduction. Results of the \ion{H}{i} observations are shown in
Sect.~\ref{section:results}. In Sect.~\ref{section:analysis} we analyze
the \ion{H}{i} data by combining them with blue magnitudes, comparing them
with
similar data on galaxies from other samples, and discussing the results.
In Sect.~\ref{section:conclusion} we present some preliminary conclusions.

\begin{table*}[h]
\centering{
\caption{\label{Tab1} Optical parameters of observed galaxies }
\begin{tabular}{rlllrrlllrl} \\ \hline \\[0.4cm]
\multicolumn{1}{c}{ \# }             &
\multicolumn{1}{c}{IAU Name}         &
\multicolumn{1}{c}{$\alpha\,(1950)$} &
\multicolumn{1}{c}{$\delta\,(1950)$} &
\multicolumn{1}{c}{V$_\mathrm{Hel}$$^{a}$}    &
\multicolumn{1}{c}{Ref.}             &
\multicolumn{1}{c}{$B_\mathrm{T}$}          &
\multicolumn{1}{c}{Ref.}             &
\multicolumn{1}{r}{$M_\mathrm{B}$}            &
\multicolumn{1}{l}{$D_\mathrm{NN}$}         &
\multicolumn{1}{l}{Other name and note}  \\
\multicolumn{1}{c}{    }             &
\multicolumn{1}{c}{        }         &
\multicolumn{1}{c}{                } &
\multicolumn{1}{c}{                } &
\multicolumn{1}{c}{\kms         }    &
\multicolumn{1}{c}{vel.}             &
\multicolumn{1}{c}{mag    }          &
\multicolumn{1}{c}{mag }             &
\multicolumn{1}{r}{mag  }            &
\multicolumn{1}{l}{Mpc     }         &
\multicolumn{1}{l}{                   }  \\
\\[0.1cm] &
\multicolumn{1}{c}{1}              &
\multicolumn{1}{c}{2}              &
\multicolumn{1}{c}{3}              &
\multicolumn{1}{c}{4}              &
\multicolumn{1}{c}{5}              &
\multicolumn{1}{c}{6}              &
\multicolumn{1}{c}{7}              &
\multicolumn{1}{l}{~~~8}              &
\multicolumn{1}{l}{~~9}              &
\multicolumn{1}{l}{10}             \\  \hline
\qq& 0750+603A   & 07$^{h}$ 50$^{m}$ 52.4 & +60$^{\circ}$ 18${'}$ 54${''}$ & 11049 & 1 &17.57 & 10 &--18.27 &   10.1& Pair of BCGs \\  
\qq& 0750+603B   & 07 \ \,50 \ \ 55.0 & +60 \,\,19 \,\,28 & 10841 & 1 &17.40 & 10 &--18.44 &   10.1& Pair of BCGs \\  
\qq& 0834+362    & 08 \ \,34 \ \ 01.6 & +36 \,\,14 \,\,36 &  9921 & 2 &17.81 & 11 &--17.82 &  12.7 & CG 212       \\  
\qq& 0847+612    & 08 \ \,47 \ \ 25.3 & +61 \,\,12 \,\,29 &  4139 & 3 &16.15 & 11 &--17.69 &   6.6 & MK~99         \\ 
\qq& 0912+599    & 09 \ \,12 \ \ 53.5 & +59 \,\,58 \,\,53 &  4150 & 3 &15.56 & 10 &--18.30 &   5.6 & MK~19         \\ 
\qq& 0919+364    & 09 \ \,19 \ \ 29.6 & +36 \,\,26 \,\,24 &  9409 & 2 &16.64 & 12 &--18.88 &   8.3 & CG~257        \\ 
\qq& 0926+606A   & 09 \ \,26 \ \ 20.1 & +60 \,\,40 \,\,02 &  4100 & 4 &16.77 & 10 &--17.02 &   6.1 & Pair of BCGs  \\ 
\qq& 0926+606B   & 09 \ \,26 \ \ 22.6 & +60 \,\,41 \,\,15 &  4190 & 4 &17.31 & 10 &--16.52 &   6.1 & Pair of BCGs  \\ 
\qq& 0938+611    & 09 \ \,38 \ \ 54.1 & +61 \,\,06 \,\,17 &  7978 & 1 &16.42 & 12 &--18.78 &   5.3 & MK~1421       \\ 
\qq& 1032+496    & 10 \ \,32 \ \ 06.4 & +49 \,\,37 \,\,14 &  8604 & 3 &17.38 & 12 &--17.96 &   5.4 &              \\  
\qq& 1044+306    & 10 \ \,44 \ \ 37.3 & +30 \,\,37 \,\,48 &  8611 & 6 &16.19 & 12 &--19.20 &   7.8 & CG~68  \\ 
\qq& 1044+307    & 10 \ \,44 \ \ 52.6 & +30 \,\,42 \,\,04 &  8466 & 6 &18.20 & 12 &--17.14 &   7.8 & CG~69   \\ 
\qq& 1048+334    & 10 \ \,48 \ \ 52.8 & +33 \,\,26 \,\,24 &  9274 & 2 &17.31 & 12 &--18.14 &   7.1 & CG~791       \\  
\qq& 1120+586A   & 11 \ \,20 \ \ 49.0 & +58 \,\,39 \,\,09 & 11131 & 1 &18.66 & 11 &--17.25 &  12.5 & Pair of BCGs   \\ 
\qq& 1120+586B   & 11 \ \,20 \ \ 56.2 & +58 \,\,38 \,\,43 & 11109 & 7 &18.36 & 11 &--17.55 &  12.5 & Pair of BCGs   \\ 
\qq& 1122+610    & 11 \ \,22 \ \ 22.6 & +61 \,\,03 \,\,28 &  9735 & 3 &17.99 & 12 &--17.63 &    5.8&              \\ 
\qq& 1124+610    & 11 \ \,24 \ \ 30.6 & +61 \,\,01 \,\,25 &  9710 & 3 &17.16 & 11 &--18.45 &    5.6&              \\ 
\qq& 1219+150$*$ & 12 \ \,19 \ \ 52.0 & +15 \,\,01 \,\,27 &  4137 & 8 &16.98 & 13 &--16.84 &    5.8 & VPC~208      \\ 
\qq& 1221+602    & 12 \ \,21 \ \ 00.4 & +60 \,\,13 \,\,07 &  6987 & 3 &16.87 & 11 &--18.09 &   11.9&              \\ 
\qq& 1225+571    & 12 \ \,25 \ \ 38.6 & +57 \,\,10 \,\,32 &  8180 & 1 &17.21 & 11 &--18.04 &    6.0&              \\ 
\qq& 1229+578    & 12 \ \,29 \ \ 24.8 & +57 \,\,48 \,\,41 &  7361 & 3 &16.72 & 11 &--18.32 &    8.7&              \\ 
\qq& 1236+122$*$ & 12 \ \,36 \ \ 51.0 & +12 \,\,28 \,\,12 &  3604 & 8 &18.18 & 14 &--15.38 &    6.2 & 8~Zw~202    \\ 
\qq& 1332+599    & 13 \ \,32 \ \ 56.8 & +59 \,\,59 \,\,53 &  9043 & 3 &16.96 & 12 &--18.51 &    5.5&              \\ 
\qq& 1353+597    & 13 \ \,53 \ \ 56.4 & +59 \,\,45 \,\,21 &  6571 & 3 &17.16 & 12 &--17.68 &    5.7& not BCG     \\  
\qq& 1354+580    & 13 \ \,54 \ \ 42.4 & +58 \,\,00 \,\,25 &  8321 & 3 &15.88 & 12 &--19.42 &    8.1&              \\  
\qq& 1408+558    & 14 \ \,08 \ \ 05.7 & +55 \,\,50 \,\,06 &  7973 & 3 &16.20 & 12 &--18.96 &    6.0&              \\  
\qq& 1420+544    & 14 \ \,20 \ \ 59.1 & +54 \,\,27 \,\,45 &  6220 & 3 &18.47 & 11 &--16.23 &   10.1&              \\  
\qq& 1427+337    & 14 \ \,27 \ \ 25.9 & +33 \,\,43 \,\,54 &  8017 & 1 &18.11 & 11 &--17.09 &    6.7& CG~447       \\  
\qq& 1541+515    & 15 \ \,41 \ \ 38.9 & +51 \,\,35 \,\,14 & 10577 & 1 &17.71 & 12 &--18.07 &    9.2&              \\  
\hline \\ 
\qq& 0745+601A   & 07 \ \,45 \ \ 44.1 & +60 \,\,08 \,\,31 &  9966 & 1 &18.21 & 10 &--17.44 &   3.6 &              \\ 
\qq& 0813+521    & 08 \ \,13 \ \ 52.7 & +52 \,\,11 \,\,56 &  7225 & 4 &17.16 & 10 &--17.77 &   4.8 &              \\  
\qq& 0943+561A   & 09 \ \,43 \ \ 17.8 & +56 \,\,10 \,\,58 &  8850 & 5 &19.0  &\ 5 &--16.42 &   5.1 &              \\  
\qq& 1040+560    & 10 \ \,40 \ \ 45.0 & +56 \,\,01 \,\,26 &  7740 & 3 &15.49 & 12 &--19.64 &   0.1 &              \\  
\qq& 1050+372A   & 10 \ \,50 \ \ 13.9 & +37 \,\,14 \,\,17 &  7669 & 9 &18.29 & 12 &--16.83 &   1.3 & CG~793 Pair of BCGs   \\ 
\qq& 1050+372B   & 10 \ \,50 \ \ 14.7 & +37 \,\,14 \,\,24 &  7711 & 2 &18.20 & 12 &--16.93 &   1.3 & CG~794 Pair of BCGs   \\ 
\qq& 1249+493    & 12 \ \,49 \ \ 35.7 & +49 \,\,19 \,\,45 &  7330 & 3 &18.11 & 11 &--16.93 &    4.0&              \\ 
\qq& 1305+547    & 13 \ \,05 \ \ 21.6 & +54 \,\,42 \,\,51 &  9714 & 3 &15.85 & 12 &--19.78 &    1.7&              \\ 
\qq& 1312+550    & 13 \ \,12 \ \ 32.8 & +55 \,\,03 \,\,45 &  9623 & 3 &16.6: & 12 &--19.02 &    2.2& Mkn~247      \\ 
\qq& 1457+540    & 14 \ \,57 \ \ 11.8 & +54 \,\,03 \,\,24 &  7970 & 3 &17.12 & 10 &--18.07 &    2.8&              \\  
\qq& 1519+496    & 15 \ \,19 \ \ 16.8 & +49 \,\,41 \,\,34 &  4548 & 3 &16.19 & 12 &--17.79 &    4.6&              \\  

\hline \\[0.3cm]
\multicolumn{10}{l}{ $^{*}$ Two BCGs from the void behind Virgo cluster   } \\
\multicolumn{10}{l}{ $^{a}$ ~Heliocentric velocities } \\
\multicolumn{10}{l}{ 1. Unpublished data from 6m telescope, and Pustilnik et al. \cite{Pustilnik02}, in preparation} \\
\multicolumn{10}{l}{ 2. Ugryumov et al. (\cite{Ugryumov98}); 3. Pustilnik et al. (\cite{Pustilnik95}); 4. Izotov et al. (\cite{Izotov93a})} \\
\multicolumn{10}{l}{ 5. Izotov et al. (\cite{Izotov97}); 6. Salzer et al. (\cite{Salzer95}); 7. Pustilnik et al. (\cite{PKLU01})} \\
\multicolumn{10}{l}{ 8. Drinkwater et al.~(\cite{Drinkwater96}); 9. Kniazev et al. (\cite{Kniazev00}) } \\
\multicolumn{10}{l}{10. Loiano 1.56m telescope data (Kniazev et al.~\cite{Kniazev02}, in preparation)} \\
\multicolumn{10}{l}{11. KPNO 0.9m telescope data (Lipovetsky et al.~\cite{Lipovetsky02}, in preparation)}   \\
\multicolumn{10}{l}{12. FLWO 1.2m telescope data (Lipovetsky et al.~\cite{Lipovetsky02}, in preparation)}  \\
\multicolumn{10}{l}{13. Young \& Currie (\cite{Young98}); 14. Zwicky et al. (\cite{Zwicky75}) } \\
\end{tabular}
}
\end{table*}

\section{The sample}
\label{section:sample}

\subsection{Sample description}

The BCGs (\ion{H}{ii} galaxies) selected for this study were chosen from
samples in the zones of SBS (Second Byurakan Survey):
R.A. = 7$^h$40$^m$ to 17$^h$20$^m$, Dec. = +49$^{\circ}$ to
+61$^{\circ}$ (Izotov et al. \cite{Izotov93a,Izotov93b}; Pustilnik et
al.~\cite{Pustilnik95}) and Case survey:
R.A. = 8$^h$00$^m$ to 16$^h$10$^m$, Dec. = +29$^{\circ}$ to +38$^{\circ}$.
The latter sample incorporates data from Augarde et al. (\cite{Augarde87}),
Weistrop \& Downes (\cite{Weistrop88,Weistrop91}), Salzer et
al.~(\cite{Salzer95}), Ugryumov (\cite{Ugryumov97}), and Ugryumov et al.
(\cite{Ugryumov98}).

To separate the most isolated BCGs from bright galaxies we used the
most complete compilation of radial velocities of galaxies from the Zwicky
catalog -- UZC by Falco et al. (\cite{Falco99}, and references therein).
To quantify our selection criterion we separated all BCGs from the indicated
samples which have no bright (that is $L > L^*$, $L^*$ corresponding to
$M_\mathrm{B}$ = --19\fm5) neighbouring galaxy closer than 5.3 Mpc.
This distance corresponds to $\sim$ 80\% of the radius of the smallest
voids identified by Kauffmann \& Fairall (\cite{Kauffmann91}).
Note, that almost all BCGs are low-mass galaxies and are substantially
fainter than $L*$. Therefore, galaxies of comparable and lower size found in
the
vicinity of many void BCGs (see Notes to  Table 1 on individual objects) do
not contradict the above definition of void BCGs. They rather indicate the
existence of a mass hierarchy in the LSS.

Eleven additional galaxies were observed with more relaxed criteria,
since they were primarily selected on an earlier version of the bright
galaxy redshift catalog, and were later discovered to have a neighbour
closer than this limit.

Galaxies from both the SBS and Case samples are selected from the
objective-prism surveys. The objects which were selected have thus
(a) strong emission lines, (b) high surface brightness.
This is relevant in the context of a comparison with published samples,
one of which is selected according to different criteria.

In the upper part of Table~\ref{Tab1} we list 28 such galaxies (plus one
non-BCG) with their coordinates
(Cols. 2 and 3), optical heliocentric velocities (Col. 4), with respective
reference (Col. 5), total blue
magnitudes (Col. 6), followed by the reference to its source (Col. 7,
see below).
The errors in radial velocity range from 20 to 100 \kms\
while the blue magnitudes have errors smaller than 0$\fm$1.
In Col. 8 we give the absolute blue magnitude, derived from $B_\mathrm{T}$
and the distance in Col. 9.
To derive the distances, the radial velocities were corrected for the solar
motion relative to Local Group (LG) centroid, according to NED (Karachentsev
\& Makarov \cite{Kara96}) and the motion of LG to Virgo cluster with a
peculiar velocity of 250~\kms (Huchra \cite{Huchra88}; Klypin et al.
\cite{Klypin01}).
For galaxies detected in \ion{H}{i}, the heliocentric velocity was adopted
from $V(\ion{H}{i})$ in Table~\ref{Tab2}, while for undetected BCGs, the
optical velocity was adopted.
No correction was done for Galaxy extinction, since this is small
at the high galactic latitudes of our sample BCGs.
In Col. 9 we give the estimate of the distance to the nearest neighbouring
bright galaxy, mainly based on the UZC catalog of Falco et al.
(\cite{Falco99}).
In Col. 10 we give alternative names for each galaxy and some short notes.
More detailed notes are given in the text.
No alternative names means that the galaxy has the prefix `SBS'.
The galaxies below the dividing line are from the complementary sample,
selected with a more relaxed criterion on $D_\mathrm{NN}$, as explained above. The
same relates to Table~\ref{Tab2}. These galaxies were not used in the
statistical analysis below.

Notes to  Table~\ref{Tab1} on individual objects: \\
{\bf 0750+603AB} -- a pair of BCGs with projected distance of 26~kpc.
  Parameters connected with \ion{H}{i} are determined assuming that the two
  galaxies  have equal  $M(\ion{H}{i})$/$L_\mathrm{B}$ values. \\
{\bf 0847+612} -- elongated central part with highly disturbed external
  morphology: bent tail at SW edge, or possibly a distinct component in an
  advanced stage of merger. \\
{\bf 0919+364} --  elongated galaxy with disturbed external part at NE, and
  in contact (15\arcsec\ to E) with $\sim$3 mag fainter and highly irregular
  galaxy.  Possibly a very advanced merger on the blue DSS-2 image.  \\
{\bf 0926+606AB} -- a pair of galaxies at a projected distance of 22~kpc.
  Although both objects fall within the same NRT (Nan\c {c}ay Radio Telescope,
  see section~\ref{section:obs}) beam, the \ion{H}{i} profile
  indicates the presence of two separate components. Decomposition of the
  \ion{H}{i}  profile was performed. \\
{\bf 0938+611} -- on the blue DSS-2 shows an irregular periphery with weak
   extension to SW (22\arcsec) -- faint companion, $\sim$5 mag. fainter. \\
{\bf 0943+561A} --  merger morphology on DSS-2 red image. \\
{\bf 1032+496} --  galaxy with disturbed external morphology. \\
{\bf 1044+306} --  highly disturbed periphery, with a probable faint companion
  $\sim$32\arcsec\ to the North. \\
{\bf 1044+307} -- the galaxy CG~69 had a position error of
   $\sim$1.9$^{\prime}$ in Pesch \& Sanduleak (\cite{Pesch83}) (cf. J.Salzer,
  private communication). The position of this galaxy in Table~\ref{Tab1} was
  measured on the Digitized Sky Survey-II (DSS) with an r.m.s. uncertainty of
  $\sim$1\arcsec. Although the NRT observations of this BCG  were conducted
  with wrong coordinates (R.A. offset of 29\arcsec, and declination offset
  of 1.9$^{\prime}$), the overall correction for the flux measurement is
  only 7\% because of the large NRT vertical beam (FWHM=22$^{\prime}$).
  This BCG is
  situated in the same void as CG~68 (1044+306), however their projected
  separation ($\sim$580 kpc) is too large to consider them a physical pair.
   The BCG is very compact ($\sim$6 kpc), with irregular external morphology,
   and with a probable irregular neighbour $\sim$1\fm0 fainter at 53\arcsec\
   ($\sim$30 kpc in projection) to the North.  \\
{\bf 1048+334} -- disturbed external morphology. Very close faint galaxy
     (12\arcsec\ to SE), almost in contact with the BCG. \\
{\bf 1050+372AB} -- a pair of BCGs in contact (projected distance of
     12\arcsec\ or $\sim$6~kpc). \\
{\bf 1120+586AB} -- a pair of BCGs with projected separation of 48~kpc and
   $\Delta V$=15~\kms.
  The galaxy SBS~1120+586B, with a previously unknown redshift,
  is very close on the sky to BCG SBS~1120+586A. It has been shown recently
(Pustilnik et al.~\cite{PKLU01}) to be a new BCG -- physical companion of the
latter BCG. Although the pointing of 100m Effelsberg radio telescope was on
the
position of SBS~1120+586A, both galaxies were well within the radiotelescope
beam, and both are considered ``detected''.
 The  B-component has a weaker emission-line spectrum, with
   $EW$([\ion{O}{iii}]5007) $\sim$25\AA. $M(\ion{H}{i})$ and
   $M(\ion{H}{i})$/$L_\mathrm{B}$ are determined for each component assuming equal
   $M(\ion{H}{i})$/$L_\mathrm{B}$ ratios. \\
{\bf 1219+150} -- both central and external parts appear irregular, probably
   a merger.\\
{\bf 1221+602} -- external parts are highly irregular, disturbed. Probable
  faint companion galaxy  $\sim$21\arcsec\ to the West. \\
{\bf 1225+571} -- approximately regular, with somewhat disturbed external
   parts. \\
{\bf 1229+578} -- disturbed external parts, possible faint satellite
  $\sim$20\arcsec\ to the West. \\
{\bf 1236+122} -- galaxy with tail to the South. Disturbed morphology. \\
{\bf 1332+599} -- disturbed external morphology. Fainter galaxy in contact
   $\sim$15\arcsec\ to South. Probable merger.\\
{\bf 1353+597} -- almost edge-on disk with warped ends and small companion
   galaxy, $\sim$3 mag. fainter,  33\arcsec\ to NW.\\
{\bf 1354+580} -- classical advanced merger. \\
{\bf 1408+558} -- large disk with highly disturbed periphery. Several faint
   irregular galaxies 25\arcsec--40\arcsec\ to NE and N: probable companions.
   The nearest one is almost in contact.  \\
{\bf 1420+544} -- compact with disturbed periphery and tail to the North. \\
{\bf 1427+337} -- on blue DSS-2: faint curved tail on the South, and probable
       faint companion $\sim$12\arcsec\ to South. \\
{\bf 1541+515} -- on blue DSS-2: disturbed periphery and faint companion in
	  contact near the western edge.

Inspection of the listed parameters shows that about half the objects
are brighter than the  $M_B$ limit (--18\fm0) usually separating dwarf
from normal galaxies, but only two objects are brighter than
$M_\mathrm{B}$=--19\fm0. Accounting for the known brightening of BCGs during
a SF burst,
relative to the luminosity in the non-active state up to 1\fm5 (on average,
0\fm75, Papaderos et al.~\cite{Papa96}), this implies that we deal
entirely with a population of low-mass galaxies that is, with gas-rich
galaxies whose masses are significantly lower than those of $L^{*}$ galaxies.
Table~\ref{Tab1} includes also three pairs of BCGs which are located in
voids; although each member of a pair is not isolated, the pairs
themselves are, and their study fits well with our intention to investigate
galaxies in very low density environments. Moreover, the notes section shows
that many objects seem to have very faint companions.

\subsection{Blue magnitudes}

Total blue magnitudes are mainly obtained through CCD photometry with three
telescopes: 1.68m Loiano telescope (Bologna University, Italy), 0.9m KPNO
telescope (USA) and 1.2m Whipple Observatory (FLWO, Arizona, USA)
during 1990-1996. These observations were conducted as part of a study
of the properties of a large BCG sample extracted from the zones of the
Second Byurakan (SBS) and Case surveys. These results will be published
elsewhere (Kniazev et al. \cite{Kniazev02}, Lipovetsky et
al.~\cite{Lipovetsky02}). Here we only
use one integrated parameter for each BCG of interest, the integrated blue
magnitude $B_\mathrm{T}$.

\subsection{Galaxy density and the distance to the nearest bright neighbour}

In order to compare the properties of void BCGs with those of BCGs
residing in a
more common environment, we need to estimate the background galaxy
density, and to check how some of the properties vary with changes in the
galaxy
density. One of such parameters is $D_\mathrm{NN}$ -- the distance to the nearest
$L^{*}$ ($M_\mathrm{B} \leq$--19\fm5) galaxy. The denser the environment, the
smaller is the $D_\mathrm{NN}$. In section~\ref{section:analysis} we illustrate
the
difference in galaxy density with  $D_\mathrm{NN}$ histograms for each of the
samples used in the comparison (Fig.~\ref{fig:Fig3}).

\section{Observations and reduction}
\label{section:obs}

Observations of 38 selected SBS and Case BCGs were conducted during the
period April 1993 to November 1996  with the
Nan\c {c}ay\footnote{The Nan\c {c}ay Radioastronomy Station is part of
Paris Observatory and is operated by the Minist\`ere de l'Education Nationale
and Institute des Sciences de l'Univers of the Centre National de la Recherche
Scientifique.}
300$\times$35m radiotelescope (hereafter NRT) and with the Effelsberg\footnote{
The Effelsberg 100m radio telescope is part of the Max-Planck-Institut
f\"{u}r Radioastronomie in Bonn} 100m
radiotelescope (hereafter the 100m telescope). 31 BCGs were observed with
the NRT and 16 BCGs were observed
with the 100m radiotelescope, with seven BCGs observed by both
telescopes. Two more BCGs, in a void behind the Virgo cluster, were
observed with the 100m radiotelescope.

\subsection{Nan\c {c}ay radio telescope data}

\ion{H}{i}-observations with the NRT are characterized by a half-power beam
width (HPBW) of
3.7$^{\prime}$~(East-West) $\times$ 22$^{\prime}$~(North-South) at
declination $\delta$=0$^\circ$. We used a dual-polarization receiver with
a system temperature of $\approx$40~K in the horizontal linear polarization
and $\approx$60~K in the vertical linear polarization. Since all observed
BCGs had known optical redshifts, we split the 1024-channel autocorrelator
into two halves, each with a bandwidth of 6.4 MHz and centered at the
frequency corresponding to the optical redshift. In this configuration, each
segment covered a velocity range of 1350~\kms. The channel spacing
was 2.6~\kms\ before smoothing and the effective resolution,
after averaging pairs of adjacent channels and Hanning smoothing, was
$\approx$21~\kms. The gain of the telescope was 1.1 K/Jy at
declination $\delta$=0$^\circ$. The observations were made in total power
(position switching) mode with 2-minute on-source and 2 minute off-source
integrations. Typically, we aimed to achieve an r.m.s. noise of 2.5
mJy per channel after smoothing. This
led to a typical integration time of 1 hour on the galaxy and one hour on
the comparison field. For galaxies with the lowest \ion{H}{i} flux densities,
the on-integration time could go up to 3--4 hours.
A noise diode, whose power was regularly monitored through observations of
known continuum and line sources, was used for the flux calibration.
Comparisons of our measured fluxes with
independent measurements of the same objects with other telescopes indicates
a consistency of the flux scale within 10\%.
The Nan\c {c}ay data was reduced using the software developed by the NRT
staff.  The two polarizations, independently detected, were averaged for each
integration to improve the sensitivity. The baselines were generally
well-fitted by a third order  or lower polynomial  and were subtracted out.

\subsection{Effelsberg 100m radiotelescope data}

The half-power beam width of the 100-m radiotelescope at Effelsberg is
9.3$'$ at a wavelength of 21-cm. We used a cooled two-channel (dual
polarisation) HEMT receiver with a system temperature of 30 K followed by
a 1024 channel autocorrelator which was split into four banks of 256
channels each. A total bandwidth of 6.25 MHz provided sufficient baseline
range and velocity coverage, as the radial velocities of all BCGs were known.
The resulting channel separation was 24.4 kHz corresponding to a velocity
resolution of 6.25~\kms\, or 10.2~\kms\ after Hanning smoothing.
The telescope gain was 1.5 K/Jy for the elevation range 20 to 80 degrees.
Observations were performed in total-power mode (position switching),
integrating for five minutes on an empty (comparison) field ahead
of source followed by five minutes on the source.  Subtracting the
off-position from the on-position reduces instrumental effects
(such as baseline problems) to approximately 10\% of their original values.
Position checks
and calibration measurements were done regularly by observing well-known
continuum and line sources. The two polarizations detected independently
were averaged for each integration to improve sensitivity.
The {\it Toolbox} software of the MPIfR was used for the data reduction.
Only modest corrections, using polynomials of
first to third order, were applied in the baseline correction procedure.

\begin{figure*}[hbtp]
 {\centering
\psfig{figure=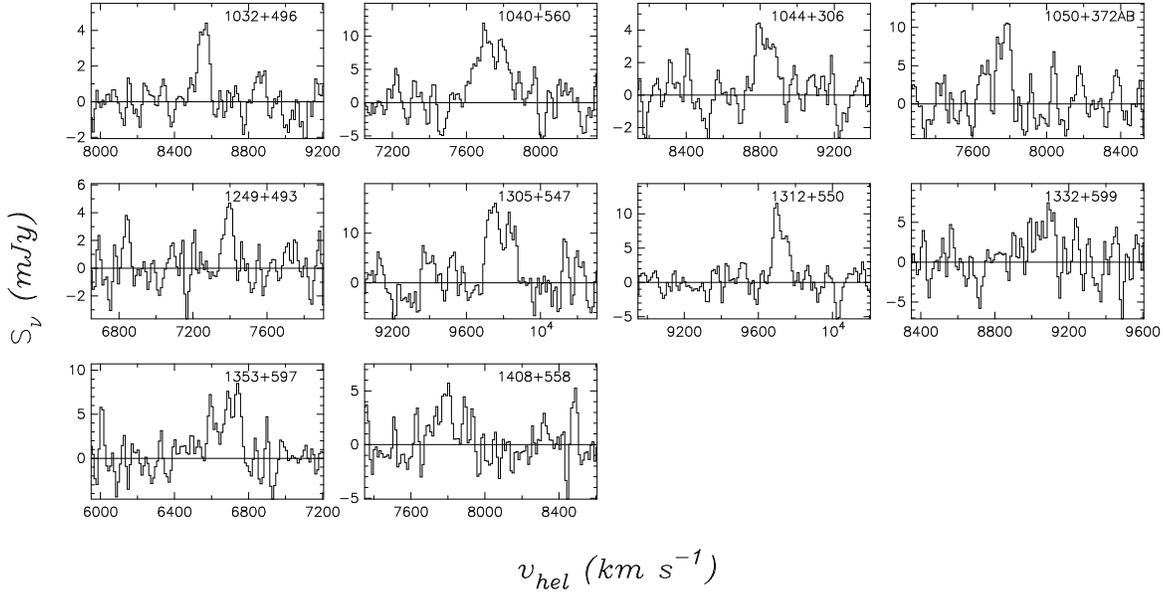,angle=-90,width=16.0cm,clip=,bbllx=73pt,bblly=47pt,bburx=444pt,bbury=734pt}
}
  \caption{\ion{H}{i}-line profiles of  10 BCGs, detected with
the Nan\c {c}ay Radio Telescope. The profiles of the seven BCGs from Table 2,
already
presented in the paper of Thuan et al.~(\cite{Thuan99}) are not shown here.
The length of the X-axes of each profile corresponds to 1300 \kms. The
velocity resolution is 21~\kms. The flux density on Y-axes is in mJy.
     }
    \label{fig:Fig1}
\end{figure*}

\begin{figure*}[hbtp]
 {\centering
\psfig{figure=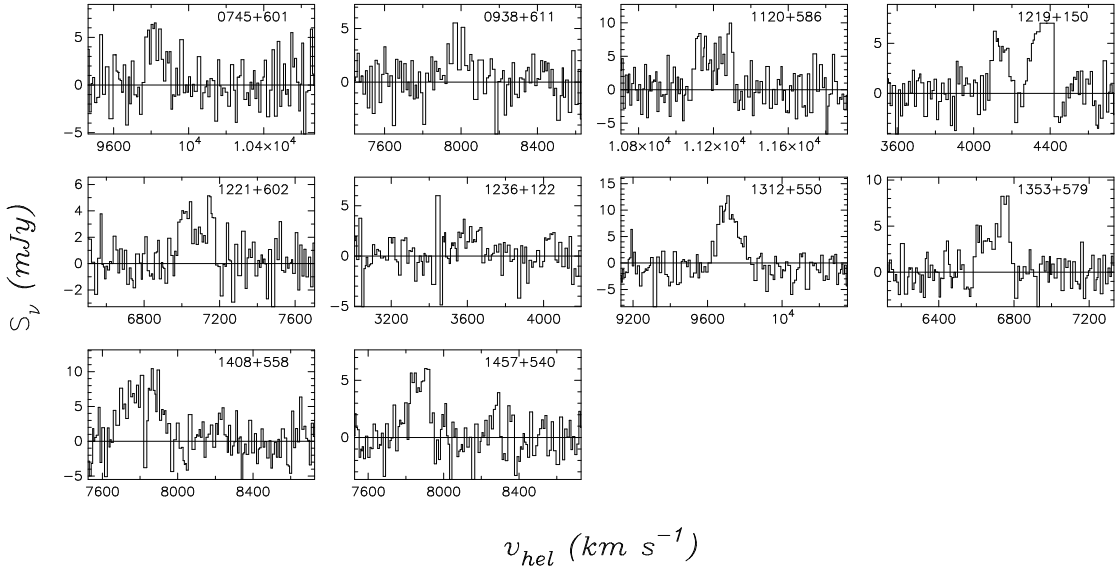,angle=-90,width=16.0cm,clip=,bbllx=56pt,bblly=43pt,bburx=436pt,bbury=748pt}
}
  \caption{\ion{H}{i}-line profiles of 10 BCGs, detected
    with the 100m Effelsberg Radio Telescope. The length of X-axes on each
    profile corresponds to 1200 \kms. Velocity resolution is 10.2~\kms.
    Flux density on Y-axes is in mJy.
     }
    \label{fig:Fig2}
\end{figure*}

\begin{table*}[h]
\centering{
\caption{\label{Tab2} \ion{H}{i} parameters of detected BCGs}
\begin{tabular}{llrrrlcrll} \hline \hline
\multicolumn{1}{c}{IAU name } & \multicolumn{1}{c}{Other} &
\multicolumn{1}{c}{$V(\ion{H}{i})$} &  \multicolumn{1}{c}{Dist.} &
\multicolumn{1}{c}{$W_\mathrm{50}$} & \multicolumn{1}{c}{$W_\mathrm{20}$} & \multicolumn{1}{c}{Obs.Flux} &
\multicolumn{1}{c}{Log} & \multicolumn{1}{l}{$M(\ion{H}{i})/L_\mathrm{B}$} & \multicolumn{1}{l}{Telescope} \\
& \multicolumn{1}{c}{name or} &
km s$^{-1}$   & Mpc  & km s$^{-1}$ &
km s$^{-1}$ & Jy km s$^{-1}$&
$M(\ion{H}{i})$  &\ & \&  \\
  & \multicolumn{1}{c}{prefix} &  & \multicolumn{1}{c}{}  &
   &   &   &
$M_{\odot}$&    &  Year    \\
\multicolumn{1}{c}{1}              &
\multicolumn{1}{c}{2}              &
\multicolumn{1}{c}{3}              &
\multicolumn{1}{c}{4}              &
\multicolumn{1}{c}{5}              &
\multicolumn{1}{c}{6}              &
\multicolumn{1}{c}{7}              &
\multicolumn{1}{l}{~~~8}           &
\multicolumn{1}{l}{~~~9}           &
\multicolumn{1}{l}{~~10}           \\ [0.1cm] \hline
0750+603AB$^{\dagger}$ & SBS    &10844$\pm$17  &146.9 &  183$\pm$34 & 238$\pm$\ 54 & 1.41$\pm$0.27 &9.86 &      & N93    \\
0750+603A              & SBS    &10844$\pm$17  &146.9 &             &              & 0.65$\pm$0.19 &9.52 & 1.05 &        \\
0750+603B              & SBS    &10844$\pm$17  &146.9 &             &              & 0.76$\pm$0.19 &9.59 & 1.05 &        \\
0847+612$^{\dagger}$   & MK~99  & 4120$\pm$12  & 58.8 &  126$\pm$23 & 179$\pm$\ 37 & 2.18$\pm$0.29 &9.24 & 0.95 & N95    \\
0912+599$^{\dagger}$   & MK~19  & 4150$\pm$13  & 59.2 &  144$\pm$26 & 206$\pm$\ 40 & 2.33$\pm$0.37 &9.27 & 0.59 & N93    \\
0926+606A$^{\dagger}$  & SBS    & 4002$\pm$24  & 59.1 &   78$\pm$23 & 120$\pm$\ 37 & 1.30$\pm$0.49 &9.00 & 1.01 & N95    \\
0926+606B$^{\dagger}$  & SBS    & 4090$\pm$24  & 59.1 &   67$\pm$23 & 120$\pm$\ 37 & 1.10$\pm$0.49 &8.93 & 1.40 & N95    \\
0938+611               & SBS    & 7980$\pm$11  &109.5 &   32$\pm$22 & ~53$\pm$\ 35 & 0.25$\pm$0.15 &8.94 & 0.17 & N94    \\
  -"-                  &        & 7983$\pm$30  &109.5 &   67$\pm$52 & 110$\pm$\ 81 & 0.38$\pm$0.16 &8.89 &      & E96    \\
1032+496               & SBS    & 8550$\pm$16  &116.7 &   74$\pm$32 & 126$\pm$\ 51 & 0.33$\pm$0.09 &9.03 & 0.45 & N94,N95\\
1044+306${*}$          &CG~68   & 8830$\pm$21  &119.5 &  127$\pm$42 & 168$\pm$\ 66 & 0.47$\pm$0.16 &9.19 & 0.21 & N95    \\
1120+586AB             & SBS    &11198$\pm$21  &152.1 &  203$\pm$11 & 219$\pm$\ 17 & 1.08$\pm$0.24 &9.77 &      & E96    \\
1120+586A              & SBS    &11198$\pm$21  &152.1 &             &              & 0.47$\pm$0.17 &9.41 & 2.06 &        \\
1120+586B              & SBS    &11198$\pm$21  &152.1 &             &              & 0.61$\pm$0.17 &9.53 & 2.06 &        \\
1219+150               &VPC~208 & 4143$\pm$29  & 58.0 &   93$\pm$13 & 112$\pm$\ 20 & 0.47$\pm$0.10 &8.56 & 0.44 & E96    \\
1221+602               & SBS    & 7083$\pm$22  & 98.1 &  191$\pm$14 & 202$\pm$\ 22 & 0.62$\pm$0.22 &9.15 & 0.53 & E96    \\
1236+122               &8~Zw~202& 3617$\pm$17  & 51.5 &  100$\pm$32 & 126$\pm$\ 50 & 0.27$\pm$0.12 &8.21 & 0.59 & E96    \\
1332+599               & SBS    & 9063$\pm$27  &124.3 &  106$\pm$54 & 148$\pm$\ 84 & 0.68$\pm$0.30 &9.22 & 0.42 & N94    \\
1353+597               & SBS    & 6680$\pm$42  & 93.0 &   95$\pm$84 & 201$\pm$131  & 1.02$\pm$0.27 &9.26 & 0.99 & N94    \\
  -"-                  &        & 6694$\pm$23  & 93.0 &   75$\pm$70 & 197$\pm$109  & 0.85$\pm$0.16 &9.26 &      & E96    \\
1408+558               & SBS    & 7812$\pm$25  &107.8 &   87$\pm$19 & 261$\pm$187  & 1.51$\pm$0.27 &9.36 & 0.38 & E96    \\
  -"-                  &        & 7779$\pm$30  &107.8 &   84$\pm$60 & 137$\pm$\ 95 & 0.36$\pm$0.22 &9.36 &      & N94    \\
\hline \\ 
0745+601               & SBS    & 9817$\pm$25  &134.7 &   60$\pm$42 & 107$\pm$\ 66 & 0.48$\pm$0.14 &9.31 & 1.40 & E96    \\
0813+521$^{\dagger}$   & SBS    & 7065$\pm$16  & 96.8 &   79$\pm$32 & 119$\pm$\ 50 & 0.83$\pm$0.26 &9.26 & 0.92 & N93    \\
1040+560               & SBS    & 7738$\pm$21  &106.3 &  211$\pm$42 & 253$\pm$\ 66 & 1.82$\pm$0.40 &9.69 & 0.43 & N94    \\
1050+372AB             &CG~793/4& 7772$\pm$10  &105.9 &   95$\pm$\ 5& ~95$\pm$ ~~5 & 0.77$\pm$0.25 &9.32 &      & N95    \\
1050+372A              &CG~793  & 7772$\pm$10  &105.9 &             &              & 0.37$\pm$0.25 &9.03 & 1.20 & N95    \\
1050+372B              &CG~794  & 7772$\pm$10  &105.9 &             &              & 0.41$\pm$0.25 &9.03 & 1.20 & N95    \\
1249+493               & SBS    & 7390$\pm$10  &101.9 &   53$\pm$20 & ~74$\pm$ ~31 & 0.27$\pm$0.13 &8.99 & 1.06 & N94,N95\\
1305+547               & SBS    & 9782$\pm$\ 7 &133.6 &  179$\pm$13 & 190$\pm$\ 21 & 2.04$\pm$0.50 &9.94 & 0.68 & N94    \\
1312+550               & SBS    & 9718$\pm$11  &132.9 &   84$\pm$22 & 116$\pm$\ 35 & 0.85$\pm$0.17 &9.72 & 0.83 & N94    \\
 -"-                   &        & 9730$\pm$24  &132.9 &  109$\pm$27 & 170$\pm$\ 42 & 1.92$\pm$0.22 &9.72 &      & E96    \\
1457+540               & SBS    & 7908$\pm$22  &109.0 &  111$\pm$\ 8& 123$\pm$\ 13 & 0.60$\pm$0.11 &9.23 & 0.64 & E96    \\
1519+496$^{\dagger}$   & SBS    & 4543$\pm$\ 4 & 64.9 &  194$\pm$\ 8& 206$\pm$\ 13 & 0.96$\pm$0.23 &8.94 & 0.44 & N94    \\
  \hline \\
\multicolumn{10}{l}{ $^{\dagger}$ Data are from Thuan et al.~(\cite{Thuan99}). } \\
\multicolumn{10}{l}{ $^{*}$ Despite the confident detection the \ion{H}{i}-velocity differs from the optical one by $\sim$200~\kms} \\
\multicolumn{10}{l}{ For BCGs detected at both telescopes the mean weighted flux is accepted and the respective $M(\ion{H}{i})$ given} \\
\multicolumn{10}{l}{ For pairs 0750+603AB, 1120+586AB and 1052+372AB with both BCGs in the beam, the first line presents the observed} \\
\multicolumn{10}{l}{ parameters. $M(\ion{H}{i})$ are calculated assuming the equal ratio $M(\ion{H}{i})/L_\mathrm{B}$ for each member of the pair} \\
\multicolumn{10}{l}{ In the system  0926+606AB the average velocity is taken for $M(\ion{H}{i})$ and $L_\mathrm{B}$} \\
\end{tabular} 
}
\end{table*}

\section{Results}
\label{section:results}

The \ion{H}{i}-profiles of the detected galaxies are shown in
Fig.~\ref{fig:Fig1} and \ref{fig:Fig2}. The profiles of seven BCGs, observed
with the NRT are not displayed in Fig.~\ref{fig:Fig1}, as they were already
shown in an earlier paper by Thuan et al. (\cite{Thuan99}). However,
their data are in Tables \ref{Tab1} and \ref{Tab2}  and included in
the analysis.
A significant part of the profiles are quite wide, but they are
not simple gaussians, or two-horned profile, as often observed for large
spirals or for ``regular'' disks. Rather, the void BCGs show complex
or asymmetric structures, indicating the presence of different dynamical
entities. Less than one-third of the profiles are narrow and single-peaked,
as typical  of dwarf galaxies.  Since the signal-to-noise
(S/N) ratio is quite poor for the majority of the presented galaxies, it is
difficult to discuss the profile shapes in further detail.

\subsection{HI-parameters for detected galaxies}

In Table~\ref{Tab2} we present those void galaxies for which we estimate a S/N
ratio of 2.5 or larger in the \ion{H}{i} detection. The Table includes the
following Cols.: (1) the IAU name for the galaxy,
(2) name prefix or alternative name,
(3) the \ion{H}{i}-velocity with its r.m.s. error (\kms),
(4) the corresponding distance in Mpc, derived from the radial velocity
the same way as in Table~\ref{Tab1},
(5) and (6) -- the widths of the \ion{H}{i} profile at 50\% and 20\% level of
the peak value,
(7) the integrated flux (in Jy~\kms),
(8) the logarithm of the total \ion{H}{i}-mass (in solar-mass units).
 For galaxies with independent data from the two telescopes we
 calculated weighted mean integrated flux, thus the \ion{H}{i}-mass entry is
 the mean from both measurements.
Col. (9) lists the value of $M(\ion{H}{i})$/$L_\mathrm{B}$ (in solar units.)
In Col. (10) we give the telescope used (N: Nan\c{c}ay, E: Effelsberg)
and  the year
of observation.
The errors of the  \ion{H}{i}-velocity, the $W_\mathrm{20}$, $W_\mathrm{50}$ and the
integrated flux
were calculated following Schneider et al. (\cite{Schneider86}).

The NRT \ion{H}{i} signal for SBS~1332+599, 0938+611 and 1408+558 is quite
low.
For the latter two objects, the 100\,m telescope detections are quite
confident,
and the data from both telescopes are consistent. The additional peak at
$V_\mathrm{hel}$ $\sim$8500~\kms\  in the spectrum of SBS 1408+558 is presumably
an artifact, since it is absent in the  spectrum obtained with the
100\,m telescope.

For BCGs pairs with close positions and velocities our detections
yield their integrated \ion{H}{i} emission. Only for SBS 0926+606A,B
was it possible to decompose the \ion{H}{i} profile, and derive the necessary
parameters separately for every component. In order to calculate their
distances we accepted their mean radial velocity, assuming their relative
velocities are either random, or in a bound system. For the remaining three
pairs we assumed that the ratio  $M(\ion{H}{i})$/$L_\mathrm{B}$ is equal for both
components, since the components are of similar type and brightness. At this
point, we calculated $M(\ion{H}{i})$ for every component of BCG pair from the
total $M(\ion{H}{i)}$ and luminosity ratio.

\subsection{Upper limits for the non-detected galaxies}

In Table~\ref{Tab3} we present upper limits to the \ion{H}{i}-flux for the
non-detected void galaxies.  The Cols. (1) to (4) are the same as in
Table~\ref{Tab2}, but the  heliocentric velocity is derived from optical
spectra, with a typical uncertainly of $\sim$100~\kms.
In Col. (5) we include the r.m.s. noise (1$\sigma$ in mJy) for
one resolution element in the velocity interval near the optical velocity of a
target galaxy.
An upper limit for the integrated \ion{H}{i} flux (in Jy~\kms) in Col. (6)
is
calculated by multiplying twice the r.m.s. value by an assumed velocity width
of 100 \kms. The respective upper limit on the logarithm of the
\ion{H}{i}-mass,
calculated in the same manner as for the galaxies in Table~\ref{Tab2}, is
given in Col. (7).  In Col. (8) we also present the upper limits on
the ratio $M(\ion{H}{i})$/$L_\mathrm{B}$.
Telescope and year of the observing run is given in Col. (9).

\begin{table*}[h]
\centering{
\caption{\label{Tab3}  Non-detections}
%
\begin{tabular}{llrrccccc} \hline \hline
\multicolumn{1}{c}{IAU name} &
\multicolumn{1}{c}{Other}    &
\multicolumn{1}{c}{\ \ $V(opt)$} &
\multicolumn{1}{c}{Dist.}      &
\multicolumn{1}{c}{r.m.s.}      &
\multicolumn{1}{c}{Up.limit} &
\multicolumn{1}{c}{Up.limit}      &
\multicolumn{1}{c}{Up.limit}      &
\multicolumn{1}{c}{Telescope} \\
\multicolumn{1}{c}{ }        &
\multicolumn{1}{c}{name or}  &
\multicolumn{1}{c}{\ \ km $s^{-1}$} &
\multicolumn{1}{c}{Mpc} &
\multicolumn{1}{c}{mJy} &
\multicolumn{1}{c}{Obs.Flux$^c$} &
\multicolumn{1}{c}{Log}   &
\multicolumn{1}{c}{$M(\ion{H}{i})/L_\mathrm{B}$}   &
\multicolumn{1}{c}{\&}  \\
\multicolumn{1}{c}{ }       &
\multicolumn{1}{c}{prefix}  &
\multicolumn{1}{c}{ }              &
\multicolumn{1}{c}{}              &
\multicolumn{1}{c}{ }              &
\multicolumn{1}{c}{Jy km $s^{-1}$} &
\multicolumn{1}{c}{$M(\ion{H}{i})$}&
\multicolumn{1}{c}{}              &
\multicolumn{1}{c}{Year}  \\
\multicolumn{1}{c}{1}              &
\multicolumn{1}{c}{2}              &
\multicolumn{1}{c}{3}              &
\multicolumn{1}{c}{4}              &
\multicolumn{1}{c}{5}              &
\multicolumn{1}{c}{6}              &
\multicolumn{1}{c}{7}              &
\multicolumn{1}{c}{8}              &
\multicolumn{1}{c}{9}              \\ [0.1cm] \hline
0834+362     &CG~212  & 9921 & 134.1 & 3.9$^a$ & 0.54 & 9.53 & 1.61 &  E96    \\
0919+364     &CG~257  & 9409 & 127.0 & 1.2$^b$ & 0.24 & 8.96 & 0.16 &  N95    \\
0943+561A    &SBS     & 8850 & 122.5 & 2.3$^b$ & 0.46 & 9.21 & 2.20 &  N94    \\
1044+307     &CG~69   & 8466 & 117.2 & 3.0$^b$ & 0.60 & 9.29 & 1.73 &  N94    \\
1048+334     &CG~791  & 9274 & 129.8 & 3.8$^b$ & 0.76 & 9.50 & 1.02 &  N95    \\
1122+610     &SBS     & 9735 & 133.4 & 1.8$^a$ & 0.25 & 9.07 & 0.59 &  E96    \\
  -"-        &        & 9735 & 133.4 & 3.4$^b$ & 0.68 &      &      &  N94    \\
1124+610     &SBS     & 9710 & 133.1 & 2.9$^b$ & 0.58 & 9.40 & 0.67 &  N94    \\
1225+571     &SBS     & 8180 & 112.9 & 2.1$^b$ & 0.42 & 9.10 & 0.49 &  N95    \\
1229+578     &SBS     & 7460 & 102.0 & 2.6$^b$ & 0.52 & 9.11 & 0.39 &  N94    \\
1354+580     &        & 8328 & 115.2 & 2.9$^a$ & 0.40 & 9.10 & 0.14 &  E96    \\ 
  -"-        &        & 8328 & 115.2 & 2.0$^b$ & 0.40 &      &      &  N94    \\
1420+544     &SBS     & 6235 &  87.4 & 1.6$^a$ & 0.22 & 8.60 & 0.81 &  E96    \\  
  -"-        &        & 6235 &  87.4 & 3.6$^b$ & 0.72 &      &      &  N94    \\
1427+337     &CG~447  & 8017 & 110.6 & 2.5$^b$ & 0.50 & 8.94 & 0.80 &  N96   \\
1541+515     &SBS     &10577 & 145.4 & 2.8$^a$ & 0.39 & 9.29 & 0.72 &  E96   \\  
    \hline \\
\multicolumn{8}{l}{$^a$ r.m.s. from effective resolution of 10.2~\kms\ of 100m telescope.} \\
\multicolumn{8}{l}{$^b$ r.m.s. from effective resolution of 21~\kms\ for NRT.} \\
\multicolumn{8}{l}{$^c$ Upper limits for integrated flux are for effective resolution of 21~\kms.} \\
\end{tabular}
}
\end{table*}

\subsection{Inter-comparison of Nan\c{c}ay and Effelsberg observations}

Tables~\ref{Tab2} and \ref{Tab3} reveal a number of objects
measured with both radio telescopes.
It is possible to use these data as an independent check of the measurement
technique. In Table~\ref{Tab2} we list four objects detected by both
instruments, while in Table~\ref{Tab3} there are three objects with upper
limits, observed  both at Nan\c{c}ay and at Effelsberg.
These seven objects form the basis of a comparison, which shows that the
measurement methods and obtained results are compatible. In
particular, we derive additional confidence from the facts that
(a) the upper limits for the undetected objects in both observatories
are similar, and
(b) the derived parameters
for the detected objects are consistent within the observational errors.

\section{Analysis and discussion}
\label{section:analysis}

\subsection{General remarks}

As already mentioned in the Introduction, our main purpose
is to study the properties of underluminous galaxies in very low density
environments. They are assumed to  have had  little interaction with other
galaxies since their formation, in comparison with similar objects in
more typical environments.
We conducted observations in the  21-cm
line of 28 BCGs in regions of very low galaxy
density  (as traced by magnitude-limited redshift surveys)  achieving a
detection rate of about 60\%. 11 BCGs in
denser environments were observed as well (lower part of Table~\ref{Tab1}).
Our intention is to investigate whether these rare
`void' BCGs (which probably comprise no more than 10--15\% of the total BCG
population) show systematical differences in comparison to the more common
BCG population.

In particular, if the main mechanism to trigger SF bursts in BCGs (and thus,
to determine their cosmological evolution) is gravitational interaction with
other galaxies, both massive and dwarfs (Pustilnik et al.~\cite{PKLU01}),
then BCGs in voids should have had much less chance to interact, and therefore
should be, on average, less evolved.
Several parameters, related to evolutionary  status,  can, in
principle, be investigated for these BCGs.  They include metallicity
of \ion{H}{ii}-regions,
colors of the underlying stellar population and (suggesting that, as a first
approximation,  evolution proceeds without significant gas exchange with the
environment) the mass fraction of \ion{H}{i} gas relative to the total baryon mass.
In order to correctly evaluate
the stellar mass, and especially that of the older population, one requires
the colors of the underlying nebulosity outside the compact region of the
current SF burst. We therefore postpone this analysis to following papers.

We address now the  evolutionary status of `void' BCGs considering only the
ratio $M(\ion{H}{i})$/$L_\mathrm{B}$. This can serve, as a ``zero-order
approximation'', as a
mean evolutionary descriptor for different BCG samples, if we accept the
following hypotheses: \\
1. The first hypothesis is that, since the BCGs in
voids and those from comparison samples were selected from objective prism
spectra independently of position or redshift within the considered ranges,
their brightening due to recent SF is similar. \\
2. Since the B-band luminosity of the underlying nebulosity measures mainly
the stellar mass formed during the last 1~Gyr, one may suggest that the SF
duty cycle in BCGs, on a timescale of $\sim$1~Gyr, depends weakly on the
type of environment.

We compare below the samples in different types of environments and
demonstrate their differences through the distribution of their
$D_\mathrm{NN}$ values -- the distance to their nearest bright galaxy.
Note that in calculating  $D_\mathrm{NN}$ here we adopted an important change in
comparison to the method used by Pustilnik et al.~(\cite{Pustilnik95}) to
analyze the large-scale distribution of BCGs.
$D_\mathrm{NN}$ was calculated there as a 3D distance, suggesting that all galaxy
velocities originate from the Hubble flow  after correcting for Virgocentric
infall.
Here we assume that sample galaxies seen close in projection ($<$ 0.5 Mpc) to
massive galaxies, and which are within $|\Delta$V $|<$ 500~\kms\
are probably physically associated.
For such pairs $D_\mathrm{NN}$ is accepted as a first approximation to be 1.5
times
the projected distance (to account for the most probable value of deprojection
correction).

\subsection{Samples of galaxies for comparison}

To have the brightening due to SF burst approximately similar in our `void'
BCGs and in the galaxies from the comparison samples, we need to use objects
with reasonably strong  emission lines. One such suitable sample of galaxies,
with a large number of members and with both $B_\mathrm{T}$ and \ion{H}{i}-fluxes
available,
is that studied by Thuan et al.~(\cite{Thuan99}) and by Pustilnik et
al.~(\cite{PKLU01}). This is a subsample of 86 SBS BCGs with $V_\mathrm{hel} <$
6000~\kms\ with $EW$([\ion{O}{iii}]5007)$>$45~\AA. In the latter work, this
sample was divided
into two groups: 26 BCGs within the borders of the Local Supercluster (LS)
and 60 remaining BCGs outside of it, that is, in the general field (GF)
(for details see Pustilnik et al.~\cite{PKLU01}).
In further comparisons we will call them respectively `LS' BCG  and `GF'
BCG samples.
To have magnitudes $M_\mathrm{B}$ of comparison SBS BCGs in Fig.
\ref{fig:Fig4}
on the same scale we recalculated their distances using the same method as
for the void BCGs. The overall changes were small -- of the order of 0\fm1
-- 0\fm3.

\begin{figure}
  \resizebox{\hsize}{!}{
  \includegraphics[angle=-90]{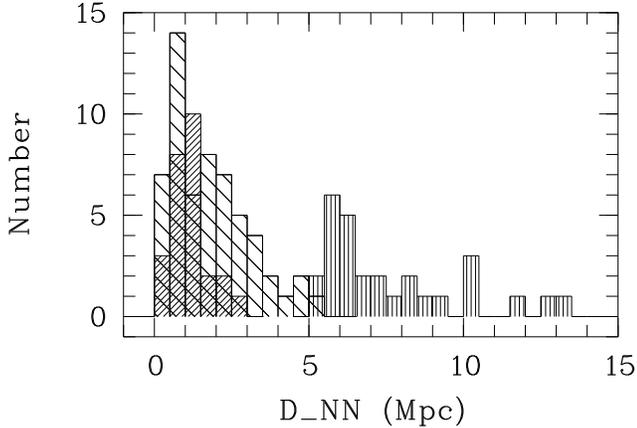}}
  \caption{Distributions of the distance to the nearest bright galaxy
  $D_\mathrm{NN}$ for 28 BCG in our void sample (vertical hatching), and two
  comparison BCG samples in SBS zone: 57 GF (general field) galaxies (sparse
  hatching, $D_\mathrm{NN} \le$ 5.5 Mpc) and 26 LS (Local Supercluster volume)
  galaxies (dense hatching, $D_\mathrm{NN} \le$ 3 Mpc).
     }
    \label{fig:Fig3}
\end{figure}

\subsubsection{General field  SBS BCG sample}
\label{GF}

The closest to the `void' BCG sample, in the aspect of the environment
galaxy density, should be that representative of the ``general field'' (GF).
The `GF' SBS subsample has both $B_\mathrm{T}$ and $M(\ion{H}{i})$ data and is,
therefore, suitable for comparison purposes. We characterize the `void'
sample by the distribution of $D_\mathrm{NN}$ (relative to bright galaxies).
To have an idea about the difference of the density of the `GF' sample,
we show its  $D_\mathrm{NN}$ distribution in Fig.~\ref{fig:Fig3} as well as
that for the `void' sample. The mean value of $D_\mathrm{NN}$
for the `GF' BCG  sample is $\sim$1.8 Mpc, while it is $\sim$7.4
Mpc for the `void' BCG sample. This implies an enhancement of a factor
of $\sim$70  in the mean galaxy
density for the GF subsample.

\subsubsection{Local Supercluster  SBS BCG sample}

A somewhat denser environment should be expected for the SBS region
within the borders of the Local Supercluster, since this includes
the outskirts of the Virgo Cluster and the poor UMa cluster. This is already
reflected in the fractions of BCGs with `massive' companions within
0.5 Mpc in both `GF' and `LS' subsamples: $\sim$31 and $\sim$54\%, as found
by Pustilnik et al.~(\cite{PKLU01}). Thus, we can use the SBS `LS' BCG sample
as well as the `GF' BCG sample to test for
differences among galaxies in regions of intermediate galaxy
density. To quantify the galaxy density of the `LS' sample  of BCGs, we show
in Fig.~\ref{fig:Fig3} its distribution of $D_\mathrm{NN}$ superposed on the
distributions for the `void' and `GF' BCG sample. Its mean, as one
would expect, is shifted to the lower value of $D_\mathrm{NN} \sim$1.1 Mpc, and
implies an $\sim$4-fold increase of the mean galaxy density, relative of the
`GF' sample, or an enhancement by almost two and a half orders of magnitude
relative to the void sample.

\begin{table*}[h]
\caption{\label{Tab4} Parameters of Virgo Cluster BCDs sample}
\begin{tabular}{rrlclcc} \\ \hline \\[0.1cm]
\multicolumn{1}{c}{VCC }             &
\multicolumn{1}{c}{$V_{Hel}$}        &
\multicolumn{1}{c}{$B_{T}$}          &
\multicolumn{1}{c}{Flux}             &
\multicolumn{1}{c}{$M_B$}             &
\multicolumn{1}{c}{$logM(\ion{H}{i})$} &
\multicolumn{1}{c}{$M(\ion{H}{i})/L_B$}  \\
\multicolumn{1}{c}{Nu.}           &
\multicolumn{1}{c}{\kms }            &
\multicolumn{1}{c}{mag}              &
\multicolumn{1}{c}{Jy~\kms }      &
\multicolumn{1}{c}{ }             &
\multicolumn{1}{c}{ }             &
\multicolumn{1}{c}{ } \\
\\[0.1cm]
\multicolumn{1}{c}{1}              &
\multicolumn{1}{c}{2}              &
\multicolumn{1}{c}{3}              &
\multicolumn{1}{c}{4}              &
\multicolumn{1}{c}{5}              &
\multicolumn{1}{c}{6}              &
\multicolumn{1}{c}{7}             \\  \hline
\\[0.1cm]
   10    &  1965 & 16.20 &  2.423 & -14.97  & 8.39  & 1.09     \\
   24    &  1280 & 16.12 &  3.538 & -15.05  & 8.55  & 1.48     \\
  144    &  2011 & 15.29 &  2.305 & -15.88  & 8.37  & 0.45     \\
  172    &  2194 & 15.38 &  4.784 & -15.79  & 8.68  & 1.01     \\
  324    &  1508 & 14.58 &  2.275 & -16.59  & 8.36  & 0.23     \\
  459    &  2103 & 15.29 &  2.465 & -15.88  & 8.40  & 0.48     \\
 1374    &  2560 & 15.00 &  1.506 & -16.17  & 8.18  & 0.22     \\
 1725    &  1051 & 15.18 &  1.893 & -15.99  & 8.28  & 0.33     \\
 1791    &  2057 & 14.85 &  6.307 & -16.32  & 8.80  & 0.82     \\
   22    &  1691 & 16.83 &  0.462 & -14.34  & 7.67  & 0.37     \\
  410    &   274 & 17.76 &  0.349 & -13.41  & 7.54  & 0.66     \\
  513    &  1828 & 15.84 &  0.270 & -15.33  & 7.43  & 0.09     \\
  562    &    45 & 16.59 &  0.348 & -14.58  & 7.54  & 0.22     \\
  985    &  1645 & 16.62 &  0.300 & -14.45  & 7.48  & 0.20     \\
 1179    &   777 & 15.98 &  0.433 & -15.19  & 7.64  & 0.16     \\
 2033    &  1508 & 15.61 &  0.410 & -15.56  & 7.61  & 0.11     \\
\hline \\[0.1cm]
\end{tabular}
\end{table*}

\subsubsection{BCDs in the Virgo cluster}

Finally, it is important to compare `void' BCGs with their
analogs in the densest environment, like that of the Virgo Cluster (VC).
Unfortunately there is no good sample of Virgo dwarfs selected by their
strong emission lines.
In the general catalog of Virgo cluster galaxies of Binggeli et
al.~(\cite{Binggeli85}, hereafter BST), there exist galaxies
classified as BCDs (blue compact dwarfs).

Based on the results of follow-up spectroscopy of some 30 VC BCDs from
the BST list, Izotov \& Guseva (\cite{Izotov89}) found that nine have
strong  [\ion{O}{iii}] emission lines, and are thus similar to the
SBS BCGs in voids considered here. The remaining 21 VC BCDs  they studied did
not show such emission lines. However, their global parameters
[$M(\ion{H}{i})$, $L_\mathrm{B}$, and size] were similar to those of the
emission-line  group of VC BCDs.
This implies that most VC BCDs represent a homogeneous group of galaxies in
various stages of recent star formation activity.  The latter conclusion
justifies (with the necessary reservations) the use of a more general VC BCD
sample for comparison  with BCGs, selected as ELGs with
\ion{H}{ii}-region type spectra.

The late-type dwarf galaxies in the BST sample have been observed in the 21 cm
line from Arecibo by Hoffman et al. (\cite{Hoffman87}). The complete sample
of 217 objects
considered by Hoffman et al. consists of all the dwarf galaxies with
$B_\mathrm{T}\leq$17.0 (with a few additions at the low surface-brightness end).
From
these we selected the objects corresponding to the morphological class of
``blue compact'', including also those of questionable classification
(BCD? in the typing of BST). This resulted in a total of 74 objects with total
blue magnitudes from the VCC and extensive \ion{H}{i} information (total flux,
line width, etc.). Objects with an \ion{H}{i} heliocentric velocity larger
than 3000~\kms\ were deleted so as to avoid contamination by
background objects. For all remaining objects it was assumed that they are
at a fixed Virgo cluster distance of 17 Mpc. The latter is
consistent with our accepted distance estimates for all considered BCGs,
if we adopt $V_\mathrm{hel}$(Virgo Cluster) = 1079~\kms\ (NED), and with the
recent
estimate of the distance to Virgo cluster from photometrical distances to
its several galaxies (Tikhonov et al. \cite{Tikhonov00}).
This does some injustice to the considerable
depth of the VC, but allows one to calculate total hydrogen contents and
luminosities in a consistent manner.

The Virgo cluster sample was selected from the Hoffman et al. list with
the following constraints: (1) the galaxies should have the morphological
mention 'BCD', or BCD and another qualifier in the VCC, (2) they should
have positive (i.e., non-zero) single-beam HI measurements, (3) the HI
velocity should be lower than 3000 km s$^{-1}$ for them to be considered
cluster members, and (4) the total HI flux should be greater than 1.5 Jy
\kms\ (for the sub-sample of high hydrogen content), or smaller
than 0.5 Jy \kms\ but still positive (for the low hydrogen content
sub-sample).

A sub-sample of $\sim$40\% of these galaxies has been studied in
detail with multi-band imaging by Almoznino \& Brosch
(\cite{Almoznino98a,Almoznino98b}). This
study revealed that only $\sim$75\% of the BCDs have H$\alpha$ emission,
and some show this as extended emission, which may not be detectable
in observations similar to those from which the ``voids'' sample was
selected. Table~\ref{Tab4} shows the relevant information for the
VC BCGs (VCC number, $V_\mathrm{hel}$, integrated $B$-band magnitude $B_\mathrm{tot}$,
 \ion{H}{i}-flux of the galaxy in Jy~\kms). The absolute magnitude $M_\mathrm{B}$,
$logM(\ion{H}{i})$ and the ratio $M(\ion{H}{i})$/$L_\mathrm{B}$, both in solar
units, are given as well.

\begin{figure*}[hbtp]
 {\centering
\psfig{figure=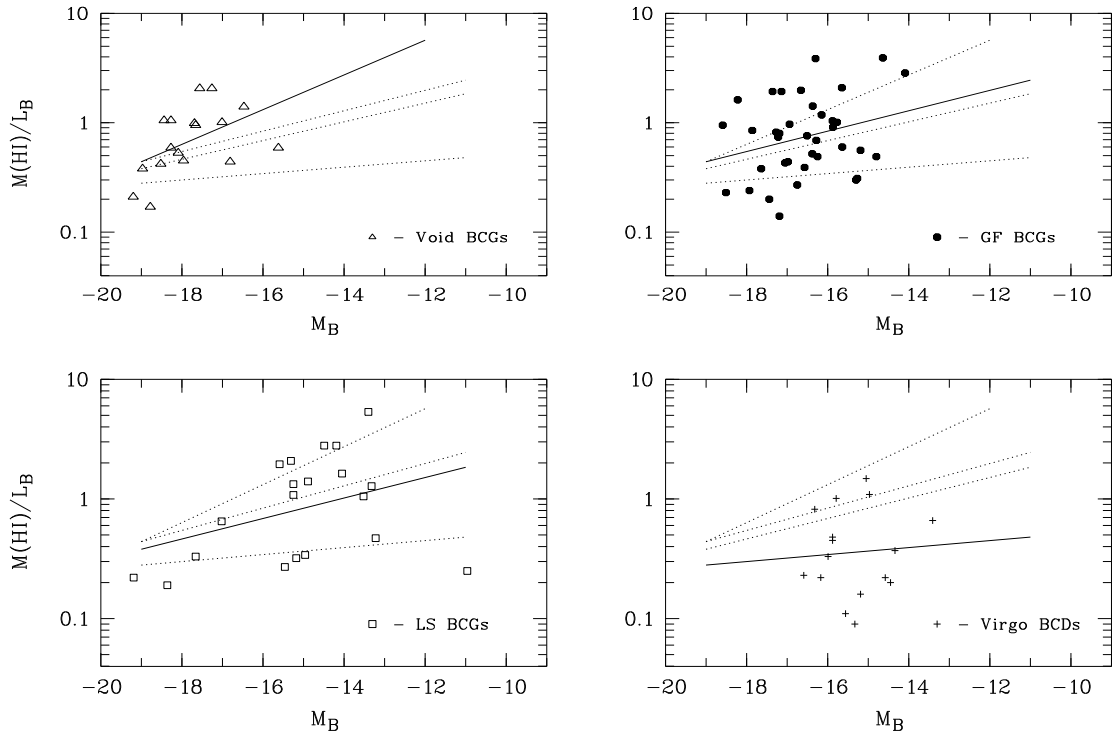,angle=0,width=21.0cm,clip=,bbllx=35pt,bblly=369pt,bburx=765pt,bbury=748pt}
}
  \caption{$M(\ion{H}{i})$/$L_\mathrm{B}$ versus $M_\mathrm{B}$ for the void BCGs and
for the three comparison samples.
   \underline{\it Top-left panel:} 17 void BCGs -- open triangles;
   \underline{\it Top-right panel:} 37 GF (general field) BCGs -- filled
   octahedrons; \underline{\it Bottom-left panel:} 20 LS (Local Supercluster
   volume) BCGs -- open squares; \underline{\it Bottom-right panel:} 16 Virgo
   cluster BCDs -- crosses.
 Solid line in each panel corresponds to the linear regression fit
 for the respective sample, as discussed in subsection \ref{section:compar}.
 Dashed lines show, in each panel, linear regression fits for the other three
   samples discussed in this subsection (top to bottom, in the order of
   Void, GF, LS and Virgo BCDs).
     }
    \label{fig:Fig4}
\end{figure*}

\subsection{Distributions of $M(\ion{H}{i})$/$L_\mathrm{B}$ for various BCG
   samples}
\label{section:compar}

Since the $M(\ion{H}{i})$/$L_\mathrm{B}$ ratio for the combined sample of
nearby galaxies increases with decreasing $L_\mathrm{B}$ (Haynes \& Giovanelli
\cite{Haynes84}; Huchtmeier \& Richter~\cite{Huchtmeier88};
Staveley-Smith et al.~\cite{Staveley92}),
we need to compare this ratio for each luminosity bin separately, that is, we
need to plot $M(\ion{H}{i})$/$L_\mathrm{B}$ versus $L_\mathrm{B}$ for each sample and
check, for example,  how the distributions differ.

In the four panels of Fig.~\ref{fig:Fig4} we show
$M(\ion{H}{i})$/$L_\mathrm{B}$ versus $M_\mathrm{B}$ with detected \ion{H}{i} flux
for all samples under consideration. They include 17
`Void' BCGs (triangles), 37 `GF' BCGs (filled octahedrons), 20 `LS' BCGs
(squares) and 16 Virgo Cluster  BCDs (crosses).

The diagrams show, as expected (excluding, perhaps, the Virgo cluster
sample), that the ratio $M(\ion{H}{i})$/$L_\mathrm{B}$ systematically increases
as the blue luminosity decreases. However, this increase is quite
different for each of the samples. The simplest way to check this is to
compare the slopes of the linear regression of
$log$($M(\ion{H}{i})$/$L_\mathrm{B}$) vs. $M_\mathrm{B}$.
Seemingly, there is a smooth transition of the slope value from the densest
environment tested here,
the VC sample, to the least dense environment represented by the
void galaxies. The slope values we measure are 0.029,
0.085, 0.093, and 0.159 for the samples in Virgo cluster, Local
Supercluster volume, General Field, and Voids, respectively.
The scatter for the Virgo BCDs is so large that one can barely suggest a
relation between the variables.

By multiplying these slopes by  --2.5, we can express them as the
more common parameter $\beta$ in the relation
$M(\ion{H}{i})$/$L_\mathrm{B} \varpropto L_\mathrm{B}^{\beta}$,
discussed, e.g., by Staveley-Smith et al.~(\cite{Staveley92}) for LSB dIs and
BCGs, and recently by Smoker et al. (\cite{Smoker00}) for emission-line
galaxies from the University of Michigan survey.
The former authors derived $\beta$=--0.3$\pm$0.1, while the latter give for
their \ion{H}{i}-detected University of Michigan survey ELGs
$\beta$=--0.2$\pm$0.1.
The values of $\beta$ for our samples of Virgo BCD and BCGs from the Local
Supercluster volume, the General Field, and from Void regions are
respectively:  --0.07$\pm$0.28, --0.21$\pm$0.12, --0.23$\pm$0.14 and
--0.40$\pm$0.18.
One can notice that the $\beta$ value for the `GF' and `LS'
BCG samples are very
close, and the regression lines approach each other.

Since the scatter of $M(\ion{H}{i})$/$L_\mathrm{B}$ in each sample
is large, and the samples are relatively small, the resulting
uncertainties in $\beta$ are also large. Therefore, despite the trend of
the slope  $\beta$  steepening with the decrease of the bright galaxy
density in BCG environment, its formal confidence level
is not high. It is important to note that, while the \ion{H}{i} integrated
flux measurement errors (up to 50\%) can result in some additional scattering
around the regression lines, the main component
of this scattering is due to other factors: the variable factor of the
$B$-band brightening due to the current SF episode can  amount
to a factor of 4--6 (1\fm5--2\fm0), and can increase the
scatter in the $M(\ion{H}{i})$/$L_\mathrm{B}$ ratio.

Comparing our results with earlier data, we conclude that the value of
$\beta$=--(0.2-0.3), derived from observations of nearby samples,
corresponds to galaxies populating the more or less typical environment --
general field, and partly, the Local Supercluster.
Void galaxies seemingly follow a steeper slope, while for BCGs in clusters,
which are on average more gas-poor, probably some additional processes take
place that dilute the luminosity dependence of $M(\ion{H}{i})$/$L_\mathrm{B}$.

The linear fits to the three BCG samples in the
$logM(\ion{H}{i})/L_\mathrm{B}$ and $M_\mathrm{B}$ plots are very close to each other
at the brightest magnitudes of the studied range,
$M_\mathrm{B}$=--19\fm0, and all three
are quite close to the fit for Virgo BCDs. However, for the fainter
luminosities the differences in average values of $M(\ion{H}{i})/L_\mathrm{B}$
among the various samples become significant. Thus, for $M_\mathrm{B}=-$17\fm0
the mean  values of $M(\ion{H}{i})/L_\mathrm{B}$ for `Void' BCGs  and VC BCDs
differ by a factor
of $\sim$3.0, as was already noticed by HHK. The difference between `Void'
and `GF' BCG for this luminosity is only $\sim$1.4, but is expected to
increase for fainter BCGs. Due to the large scattering in the data points,
the differences found between  BCGs in voids and in more common environments
should be considered tentative. Larger samples, and higher S/N
data, are necessary to further confirm and study this phenomenon.

A possible explanation could be connected with
the difference in the ability of a galaxy to retain its
hydrogen; in a void environment
this may be easier than in a denser neighbourhood, such as in a cluster.
The increase of the $M(\ion{H}{i})$/$L_\mathrm{B}$ difference in the four
galaxy samples considered here, along with the decrease of galaxy luminosity,
is consistent with
the hypothesis that low-mass galaxies lose their gas mainly due to
interactions with
other galaxies. The smaller a galaxy the more severe is the effect of
interaction with a colliding partner.

We note that the values of $M(\ion{H}{i})$/$L_\mathrm{B}$ cited for
`void' and other
BCGs are not the largest among gas-rich galaxies; some of the LSB dwarfs
studied by van Zee et al.~(\cite{vanZee95}) have a ratio several times
higher. However, accounting for the significant brightening of void BCG
progenitors during SF bursts, their preburst value of
$M(\ion{H}{i})$/$L_\mathrm{B}$
could be comparable to that of the most gas-rich Low Surface Brightness
Dwarfs.

\subsection{Interactions as SF triggers in void BCGs}

Based on the notes in Table~\ref{Tab1}, we would like to comment briefly on
the probable
tidal action of the nearby neighbours with luminosities comparable to or
lower than those of
studied BCGs. Of 28 BCGs in upper part of Table~\ref{Tab1} (SBS~1353+597 is
non-ELG) six galaxies belong to "pairs" with $\Delta$V$\le$150~\kms, and
projected distances in the range of 20--30 kpc. Six more galaxies --
0847+612, 0912+599, 0919+364, 1219+150, 1332+599 and 1354+580
show signs of merger morphologies.
A dozen more BCGs show disturbed external morphologies, that could be
taken as an indication
of recent interactions. Most have significantly fainter
galaxies ($\Delta$B=2--4 mag.) either in contact, or at the angular distances
of 15\arcsec--30\arcsec, corresponding to projected linear distances of
$\sim$(7--20) kpc. While their radial velocities are unknown, a significant
fraction of these faint "companions" appear to be physical companions of
the void BCGs under consideration. A more detailed
analysis of the possible tidal trigger of SF bursts in such pairs
was presented by
Pustilnik et al.~(\cite{PKLU01}). Taking the numbers above, we can expect that
up to $\sim$80\% of void BCGs have their SF triggered by interactions with
galaxies in their local
environment. This is consistent with the results obtained by
Pustilnik et al.~(\cite{PKLU01}) for the general field. In voids, all
these nearby galaxies, by the definition of voids, should be of relatively low
luminosity/mass.

It is worthwhile to note that the difference in the
$M(\ion{H}{i})$/$L_\mathrm{B}$ ratio  between BCGs in voids and in the general
field
discussed previously is quite small, despite the difference in
the environment density of bright galaxies of factor of 70 (see
section~\ref{GF}). If galaxy
interactions are the main factor of \ion{H}{i} gas astration and removal,
this finding indicates that the densities of lower-mass galaxies in voids
and in the general field differ much less than that of $L_{*}$
galaxies.

\section{Conclusions}
\label{section:conclusion}

1. In this paper we presented observational data on the integrated \ion{H}{i}
   emission of 28 BCGs residing in regions of very low galactic density
   (voids).
   About 60\% of them were detected, and their \ion{H}{i}-masses were
   determined and used  in further analysis.                              \\
2. The parameter $M(\ion{H}{i})$/$L_\mathrm{B}$ was derived and its dependence on
   $L_\mathrm{B}$  was analyzed for void BCGs and for some comparison BCG samples.
   Our analysis indicates a trend in the variation of the index
   $\beta$ of the power-law fit of $M(\ion{H}{i})$/$L_\mathrm{B}$ $\varpropto$
   $L_\mathrm{B}^{\beta}$ with decreasing bright galaxy density, from cluster
   to the void environment, with a full range of $\beta$ from --0.1 to --0.4. \\
3. Consistently, the mean $M(\ion{H}{i})$/$L_\mathrm{B}$ value for a given
   $L_\mathrm{B}$
   increases with the decrease of bright galaxy density. The `void' BCGs have
   the highest $M(\ion{H}{i})$/$L_\mathrm{B}$ mean values, which hints to their
   younger evolution status or lower gas loss due to interactions. \\
4. We note that a significant fraction of BCGs in voids (probably up to
   $\sim$60\%) form  either interacting pairs with low-mass galaxies of
   similar luminosity,
   or show morphological signs of interaction and/or close faint
   neighbours; in addition $\sim$20\% demonstrate merging in various stages.
   This implies, in accordance with  the results of Grogin
   \& Geller~(\cite{Grogin00b}),  that even in voids, interactions with
   the local environment of low-mass galaxies may be one of the
   main channels to trigger SF bursts in BCG progenitors.
\begin{acknowledgements}

We are grateful to the NRT Comittee de Programmes and to the Program Committee
of the Effelsberg Radiotelescope for observational time, and to T.X.Thuan
for help during the initial phase of this project. J.Salzer kindly provided
the information on the wrong position of CG~69. The notes and suggestions
of the anonymous referee
were very helpful in improving the text and the result
presentation.  SAP is grateful
for the hospitality of the staff of Observatoire de Paris-Meudon, ARPEGES,
and of the Wise Observatory of Tel Aviv University, where parts of this work
were performed. The authors  are grateful to A.Kniazev for providing his
unpublished photometry data and to A.Ugryumov for providing his code for
calculating $D_\mathrm{NN}$.
Part of this project was performed within the INTAS Research Grant 96-0500;
SAP, JMM, and NB are grateful to the INTAS foundation for this grant.
NB is grateful for the continued
support of the Austrian Friends of Tel Aviv University and acknowledges
support from a  Center of Excellence award from the Israel
Science Foundation.  The use of the Digitized Sky Survey (DSS-II) was very
helpful for checking the morphology and local environment of the studied
galaxies.
\end{acknowledgements}

\renewcommand{\baselinestretch}{0.5}





\begin{thebibliography}{99}

\bibitem[1998a]{Almoznino98a}
  Almoznino, E., \& Brosch, N. 1998a, MNRAS, 298, 920
\bibitem[1998b]{Almoznino98b}
  Almoznino, E., \& Brosch, N. 1998b, MNRAS, 298, 931
\bibitem[1987]{Augarde87}
 Augarde, R., Figon, P., Kunth, D., \& Sevre P. 1987, A\&A, 185, 4
\bibitem[1986]{Bardeen86}
 Bardeen, J.M., Bond, J.R., Kaiser, N., \& Szalay, A.S. 1986, ApJ, 304, 15
\bibitem[1985]{Binggeli85}
  Binggeli, B., Sandage, A., \& Tammann G.A. 1985, AJ, 90, 1681.
\bibitem[2001]{Bode01} Bode, P., Ostriker, J., \& Turok N. 2001,
   ApJ, 556, 93
\bibitem[1998]{Brosch98}
 Brosch, N., Almoznino, E., \& Hoffman, G.L.  1998, A\&A,  331, 873.
\bibitem[2001]{Cen01}
    Cen, R. 2001, ApJ, 546, L81
\bibitem[2000]{Cen00}
   Cen, R., \& Ostriker J. 2000, ApJ, 538, 83
\bibitem[1997]{Cruzen97}
   Cruzen, S.Y., Weistrop, D., \& Hoops S.G. 1997, AJ, 113, 1983
\bibitem[1986]{Dekel86}
 Dekel, A., \& Silk, J. 1986, ApJ, 303, 39
\bibitem[1989]{Doroshkevich89}
  Doroshkevich, A.G., Klypin, A.A., \& Khlopov, M.Y. 1989, MNRAS, 239, 923
\bibitem[1997]{Dalcanton97}
  Dalcanton, J.J., Spergel, D.N., \& Summers F.J. 1997, ApJ, 482, 659
\bibitem[1996]{Dominguez96}
  Dominguez-Tenreiro, R., Alimi, J.-M., Serna, A., \&  Thuan T.X.
  1996, ApJ, 469, 53
\bibitem[1996]{Drinkwater96}
  Drinkwater, M.J., Currie, M.J., Young, C.K., Hardy, E., \& Yearsley J.M.
  1996,  MNRAS, 279, 595
\bibitem[1999]{Falco99}
  Falco, E.E., Kurtz, M.J., Geller, M.J., et al.  1999,  PASP, 111, 438
\bibitem[2000a]{Grogin00a}
   Grogin, N., \& Geller, M. 2000a, AJ, 118, 2561
\bibitem[2000b]{Grogin00b}
  Grogin, N., \& Geller, M. 2000b, AJ, 119, 32
\bibitem[1984]{Haynes84} Haynes, M.P., \& Giovanelli, R. 1984, AJ, 89, 758
\bibitem[1987]{Hoffman87}
  Hoffman, G.L., Helou, G., Salpeter, E.E., Glosson, J., \& Sandage, A.
  1987, ApJS, 63, 247.
\bibitem[1998]{Hopp98}
  Hopp, U. 1998, in: The Evolving Universe. Selected Topics on Large-Scale
  Structure and on the Properties of Galaxies. eds. D.Hamilton et al.,
  Dordrecht, Kluwer. Series: Astrophysics and Space Science Library, Vol.
  231, p.59
\bibitem[1995]{Hopp95}
  Hopp, U., Kuhn, B., Thiele, U., et al. 1995, A\&AS, 109, 537
\bibitem[1988]{Huchra88} Huchra, J.P. 1988, in ASP Conf. Ser. 4:
   The Extragalactic  Distance Scale. p.257
\bibitem[1988]{Huchtmeier88}
  Huchtmeier, W.K., \& Richter, O.-G. 1988, A\&A, 203, 237  (HHK)
\bibitem[1997]{Huchtmeier97}
  Huchtmeier, W.K., Hopp, U., \& Kuhn, B. 1997, A\&A, 319, 67
\bibitem[1989]{Izotov89}
  Izotov, Y.I., \& Guseva, N.G., 1989, Astrofizika, 30, 564
\bibitem[1993a]{Izotov93a}
  Izotov, Y.I., Guseva, N.G., Lipovetsky, V.A., et al.
  1993a,  Astr. \& Astrophys. Transact., 3, 179
\bibitem[1993b]{Izotov93b}
 Izotov, Y.I., Lipovetsky, V.A., Guseva, N.G., Kniazev, A.Y., \& Stepanian,
  J.A.  1993b, in:
  The Feedback of Chemical Evolution on  the Stellar Content of Galaxies.
   eds. D.Alloin \& G.Stasinska, 127
\bibitem[1997]{Izotov97}
  Izotov, Y.I., Thuan, T.X., \& Lipovetsky, V.A. 1997, ApJS, 108, 1
\bibitem[1996]{Kara96}
   Karachentsev, I.D., \& Makarov, D.I. 1996, AJ, 111, 794
\bibitem[1991]{Kauffmann91}
  Kauffmann, G., \& Fairall, A. 1991, MNRAS, 248, 313
\bibitem[1993]{Klypin93}
 Klypin, A., Holtzman, J., Primack, J., \& Regos, A. 1993, ApJ, 416, 1
\bibitem[2001]{Klypin01}
 Klypin, A., Hoffman, Y., Kravtsov,A., \& Gottl\"ober, S. 2001, ApJ,
   submitted = astro-ph/0107104
\bibitem[2000]{Kniazev00} Kniazev, A.Y., Pustilnik, S.A., Ugryumov, A.V.,
   \& Kniazeva, T.F. 2000, Astronomy Letters, 26, 163
\bibitem[2002]{Kniazev02}
   Kniazev, A.Y. et al.  2002, in preparation
\bibitem[2000]{Lee00}
  Lee, J.C., Salzer, J.J., Rosenberg, J.L., \& Law, D.A. 2000, ApJ, 536, 584
\bibitem[1995]{Lindner95}
  Lindner, U., Einasto, J., Einasto, M., et al.  1995, A\&A, 301, 329
\bibitem[1996]{Lindner96}
  Lindner, U., Einasto, M., Einasto, J., et al. 1996, A\&A, 314, 1
\bibitem[2002]{Lipovetsky02} Lipovetsky, V.A., Thuan, T.X., Richter, G. et al.
  2002, in preparation
\bibitem[1996]{Ostriker96}
  Ostriker, J. 1996, ARA\&A, 31, 689
\bibitem[1996]{Papa96} Papaderos, P., Loose, H.-H., Fricke, K.J., \& Thuan,
  T.X. 1996,  A\&A, 314, 59
\bibitem[2001]{Peebles01} Peebles, P.J.E. 2001, ApJ,  557, 495
\bibitem[1983]{Pesch83} Pesch, P., \& Sanduleak, N. 1983, ApJS, 51, 171
\bibitem[1996]{Popescu96}
 Popescu, C.C., Hopp, U., Hagen, H.-J., \& Els\"asser, H. 1996, A\&AS, 116,
   43
\bibitem[1997a]{Popescu97a}
 Popescu, C.C., Hopp, U., \& Els\"asser, H. 1997a, A\&A, 325, 881
\bibitem[1997b]{Popescu97b}
 Popescu, C.C., Hopp, U., \& Els\"asser, H. 1997b, A\&A, 328, 756
\bibitem[1998]{Popescu98}
 Popescu, C.C., Hopp, U., Hagen, H.J., \& Els\"asser, H. 1998, A\&AS,
  133, 13
\bibitem[1994]{Pustilnik94} Pustilnik, S.A., Ugryumov, A.V., \& Lipovetsky,
  V.A. 1994, Astr.Astroph.Transact., 5, 75
\bibitem[1995]{Pustilnik95}
  Pustilnik, S.A., Ugryumov, A.V., Lipovetsky, V.A., Thuan, T.X., \&
  Guseva N.G.  1995, ApJ, 443, 499
\bibitem[2001]{PKLU01}
  Pustilnik, S.A., Kniazev, A.Y., Lipovetsky, V.A., \& Ugryumov, A.V.
   2001, A\&A, 373, 24
\bibitem[2002]{Pustilnik02} Pustilnik, S.A., Kniazev, A.Y., Ugryumov, A.V.,
 et al. 2002, in preparation
\bibitem[1997]{Sage97}
  Sage, L.J., Weistrop, D., Cruzen, S., \& Kompe, C. 1997, AJ, 114, 1753
\bibitem[1989]{Salzer89}
 Salzer, J.J. 1989, ApJ, 347, 152
\bibitem[1995]{Salzer95}
 Salzer, J.J., Moody, J.W., Rosenberg, J.L., Gregory, S.A., \& Newberry M.V.
  1995, AJ, 109, 2376
\bibitem[1986]{Schneider86}
  Schneider, S.E., Helou, G., Salpeter, E.E., \& Terzian, Y.  1986, AJ,
  92, 742
\bibitem[2000]{Smoker00}
   Smoker, J.V., Davies, R.D., Axon, D.J., \& Hummel, E. 2000, A\&A, 361, 19
\bibitem[1992]{Staveley92} Staveley-Smith, L., Davies, R.D., \& Kinman, T.D.
  1992,  MNRAS, 258, 334
\bibitem[1996a]{Szomoru96a}
  Szomoru, A., van Gorkom, J.H., \& Gregg, M.D. 1996a, AJ, 111, 2141
\bibitem[1996b]{Szomoru96b}
  Szomoru, A., van Gorkom, J.H., Gregg, M.D. \& Strauss, M.A. 1996b, AJ, 111,
  2150
\bibitem[1991]{Thuan91}
 Thuan, T.X., Alimi, J.-M., Gott, J.R., \& Schneider, S.E. 1991, ApJ, 370, 25
\bibitem[1992]{Thuan92} Thuan, T.X., Balkowski, C., \& Van, J.T.T., editors,
   Physics of Nearby Galaxies. Nature or Nurture ? Editions Fronti\'eres,
   Gif-sur-Yvette, 1992, 540 pp.
\bibitem[1999]{Thuan99}
  Thuan, T.X., Lipovetsky, V.A., Martin, J.-M., \& Pustilnik, S.A. 1999,
   A\&AS, 139, 1
\bibitem[2000]{Tikhonov00} Tikhonov, N.A., Galazutdinova, O.A., \&
   Drozdovsky, I.O. 2000, Astrofizika, 43, 503
\bibitem[1997]{Ugryumov97}
  Ugryumov, A.V., 1997, Ph.D.Thesis, Special Astrophysical Observatory RAS,
       Nizhnij Arkhyz
\bibitem[1998]{Ugryumov98}
  Ugryumov, A.V., Pustilnik, S.A., Lipovetsky, V.A., Izotov, Y.I., \& Richter,
   G.M. 1998, A\&AS, 131, 285
\bibitem[1995]{vanZee95} van Zee, L., Haynes, M.P., \& Giovanelli, R. 1995,
   AJ, 103, 990
\bibitem[1996]{Vennik96}
  Vennik, J., Hopp, U., Kovachev, B., Kuhn, B., \& Els\"{a}sser, H. 1996,
  A\&AS,  117, 261
\bibitem[1988]{Weistrop88}
  Weistrop, D., \& Downes, R.A. 1988, ApJ, 331, 172
\bibitem[1991]{Weistrop91}
  Weistrop, D., \& Downes, R.A. 1991, AJ, 102, 1680
\bibitem[1994]{Weistrop94}
  Weistrop, D. 1994,
  in: Violent Star Formation: from 30 Doradus to Quasars. Proc. of
  Workshop in Spain, ed. G.Tenorio-Tagle, et al., Cambridge, Cambridge
  Univ. Press, p. 100
\bibitem[1991]{White91}
  White, S., \& Frenk, C. 1991, ApJ, 379, 52
\bibitem[1998]{Young98}
  Young, C.K., \& Currie, M.J. 1998, A\&AS, 127, 367
\bibitem[1975]{Zwicky75}
  Zwicky, F., Sargent, W. L. W., \& Kowal, C. T. 1975, AJ, 80, 545
\end{thebibliography}
\end{document}